\documentclass[final,1p,times]{elsarticle}

\usepackage{graphics}

\usepackage{amssymb}
\usepackage{subeqn}
\usepackage{color}

\biboptions{sort&compress}

\journal{}

\begin{document}

\begin{frontmatter}

\title{Generalized lattice Boltzmann method: Modeling, analysis, and elements}

\author{Zhenhua Chai\fnref{a,b}}\ead{hustczh@hust.edu.cn}
\author[a,b]{Baochang Shi\corref{cor1}}\cortext[cor1]{Corresponding author at: School of Mathematics and Statistics, Huazhong University of Science and Technology, Wuhan, 430074, China. Tel./fax: +86 27 8754 3231.
\hspace*{20pt}}\ead{shibc@hust.edu.cn}
\address[a]{School of Mathematics and Statistics, Huazhong University of Science and Technology, Wuhan, 430074, China}
\address[b]{Hubei Key Laboratory of Engineering Modeling and Scientific Computing, Huazhong University of Science and Technology, Wuhan 430074, China}

\begin{abstract}
In this paper, we first present a unified framework for the modelling of generalized lattice Boltzmann method (GLBM). We then conduct a comparison of the four popular analysis methods (Chapman-Enskog analysis, Maxwell iteration, direct Taylor expansion and recurrence equations approaches) that have been used to obtain the macroscopic Navier-Stokes equations and nonlinear convection-diffusion equations from the GLBM, and show that from mathematical point of view, these four analysis methods are equivalent to each other. Finally, we give some elements that are needed in the implementation of the GLBM, and also find that some available LB models can be obtained from this GLBM.
\end{abstract}


\end{frontmatter}

%
%

\section{Introduction}
Fluid flows, heat and mass transfer are the most common basic phenomena in nature, science and engineering, and usually, they can be described by the classical Navier-Stokes equations (NSEs) and nonlinear convection-diffusion equations (CDEs) \cite{Bird2002}. However, it is difficult to obtain their analytical solutions due to the coupling between the nonlinear NSEs and CDEs. With the development of computing technique, the numerical simulation has been becoming more popular in the study of the complex fluid flows, heat and mass transfer. The lattice Boltzmann (LB) method, as one of mesoscopic numerical approaches, has attained increasing attention, and also gained a great success in the simulation of the complex physical systems for its distinct advantages in kinetic background, treatment of complex boundary conditions, and computational efficiency in parallel systems \cite{Chen1998,Succi2001,Guo2013,Kruger2017}.

Based on the collision term, the LB models developed in the past years can be classified into three different kinds. The first one is the lattice BGK model in which only one relaxation time is introduced into the collision term, hence it is also usually called the single-relaxation-time (SRT) model. Historically, this model, as one of the most popular LB models for its simplicity and efficiency, has been proposed independently by two different research groups \cite{Qian1992,Chen1992}, and later, it is also found that the SRT model can be considered as a special discretized form of the continuous BGK-Boltzmann equation \cite{He1997}. Almost at the same time, the second one named multiple-relaxation-time (MRT) or generalized LB model is developed by d¡¯Humi\`{e}res \cite{dHumieres1992}. Compared to the SRT model, the collision of MRT model is depicted with a matrix that allows for a decoupled relaxation of different moments, and includes all possible degrees of freedom that can be used to optimize the LB method in terms of stability and accuracy \cite{Lallemand2000,dHumieres2002,Pan2006,Luo2011,Chai2016a,Chai2016b}. The last one is two-relaxation-time (TRT) model, which is first developed by Ginzburg \cite{Ginzburg2005a,Ginzburg2008}. This model, through introducing two different relaxation parameters in the collision term, possesses the flexibility of the MRT model and the simplicity of SRT model, and also retains the computational efficiency \cite{Luo2011,Zhao2018}. Actually, besides the Ginzburg's TRT model \cite{Ginzburg2005a,Ginzburg2008}, the lattice kinetic (LK) scheme \cite{Inamuro2002}, the modified lattice kinetic (MLK) schemes \cite{Yang2014,Wang2015}, and the regularized LB (RLB) models \cite{Latt2006,Chen2006,WangL2015,Wang2016,WangL2018} can also be viewed as some special TRT models, as pointed out in Refs. \cite{Zhao2018,Wang2018}. Meng et al. \cite{Meng2015} also developed another version of MRT model which combines the commonly used MRT and MLK schemes \cite{dHumieres1992,Lallemand2000,Inamuro2002,Yang2014,Wang2015}, and found that compared to the popular MRT model \cite{dHumieres1992,Lallemand2000}, their model is more stable. Recently, Zhao et al. \cite{Zhao2019} proposed a block triple-relaxation-time LB model for nonlinear anisotropic CDEs in which a block diagonal relaxation matrix composed of three blocks is introduced. They also found that that above mentioned RLB models and the MLK schemes are the particular forms of the block triple-relaxation-time LB model. However, one can show the model developed by Zhao et al. \cite{Zhao2019} is also a particular MRT version.

As discussed above, although many different LB models have been developed in the past decades, all of them can be viewed as the special forms of MRT model. It should also be noted that as a more general version, however, the MRT model still has some issues needed to be addressed. For instance, it is unclear how to choose the eigenvectors of the collision matrix in the MRT model besides the ones corresponding to the conserved variables (e.g., the density and momentum appeared in the NSEs), this is mainly caused by their nonuniqueness. In addition, the analysis of MRT model is usually limited to the specified dimensional space and lattice structure (e.g., D2Q9 lattice \cite{dHumieres1992,Guo2013,Chai2016a}), and compared to the SRT model, it seems more difficult to present a general analysis to derive the macroscopic governing equations. Here we would like to point out that Adhikari and Succi have proposed a guideline to choose the eigenvectors through introducing the notion of duality \cite{Adhikari2008}. Furthermore, based on the Taylor expansion, Kaehler and Wagner have performed a general analysis (not restricted to a specific lattice structure) to the MRT model and also gave some necessary requirements on the equilibrium distribution function and the collision matrix, while in their analysis, some assumptions on the space and time derivatives are made \cite{Kaehler2013}.
On the other hand, we also note that some different analysis methods (e.g., the Chapman-Enskog analysis \cite{Guo2013,Chapman1970}, Maxwell iteration \cite{Ikenberry1956,Yong2016,Zhao2017,Zhang2019}, direct Taylor expansion or the Taylor expansion in Refs. \cite{Kaehler2013,Holdych2004,Wagner2006}, and recurrence equations \cite{dHumieres2009,Ginzburg2015} approaches) used to derive the macroscopic governing equations from the LB models are developed from different physical and mathematical points of view, while it is still unclear whether they are equivalent to each other in obtaining these macroscopic governing equations. In this paper, we will first present a unified framework for the modeling of GLBM. Then we also carry out a comparison among four different analysis methods, and following the previous work \cite{Kaehler2013}, a general analysis on the GLBM with some different methods is also performed. To implement the GLBM, some elements are also provided.

\section{Generalized lattice Boltzmann method}

The semi-discrete evolution equation of the GLBM with the D$d$Q$q$ ($q$ discrete velocities in $d$-dimensional space, $d=1\sim 3$) lattice model is written as (see Appendix A for details)
\begin{equation}\label{eq:2-1}
 f_j(\mathbf{x}+\mathbf{c}_j \Delta t,t+\Delta t)=f_j(\mathbf{x}, t)-\mathbf{\Lambda}_{jk} f_k^{ne}(\mathbf{x}, t)+\Delta t \big[G_j(\mathbf{x},t)+F_j(\mathbf{x}, t)+\frac{\Delta t}{2}\bar{D}_j F_j(\mathbf{x}, t)\big],
\end{equation}
where $f_j(\mathbf{x}, t)$ is the distribution function at position $\mathbf{x}$ in $d$-dimensional space and time $t$ along the velocity $\mathbf{c}_j$, $\mathbf{\Lambda}=(\mathbf{\Lambda}_{jk})$ is a $q \times q$ invertible collision matrix, $f_j^{ne}(\mathbf{x}, t)=f_j(\mathbf{x}, t)-f_j^{eq}(\mathbf{x}, t)$ is the non-equilibrium distribution function, and $f_j^{eq}(\mathbf{x}, t)$ is the equilibrium distribution function. $F_j(\mathbf{x}, t)$ is the distribution function of a source or forcing term, and $G_j(\mathbf{x}, t)$ is the \textit{auxiliary} source distribution function for removing additional terms. $\Delta t$ is the time step, $\bar{D}_j=\partial_t +\gamma \mathbf{c}_j\cdot \nabla$ with $\gamma \in \{0,1\}$ being a parameter to be determined later.

Generally, the evolution process based on Eq. (\ref{eq:2-1}) can be divided into two sub-steps, i.e., collision and propagation,
\begin{eqnarray}\label{eq:2-2}
\textbf{Collison:}\ \  \tilde{f}_j(\mathbf{x}, t)=f_j(\mathbf{x}, t)-\mathbf{\Lambda}_{jk} f_k^{ne}(\mathbf{x}, t)+\Delta t \big[G_j(\mathbf{x},t)+F_j(\mathbf{x}, t)+\frac{\Delta t}{2}\bar{D}_j F_j(\mathbf{x}, t)\big],  \nonumber\\
\textbf{Propagation:}\ \  f_j(\mathbf{x}+\mathbf{c}_j \Delta t,t+\Delta t)=\tilde{f}_j(\mathbf{x}, t), \ \ \ \ \ \ \ \ \ \ \ \ \ \ \ \ \ \ \ \ \ \ \ \ \ \ \ \ \ \ \ \ \ \ \ \ \ \ \ \ \ \ \ \ \ \ \ \ \ \ \ \ \ \ \ \ \ \ \ \
\end{eqnarray}
where $\tilde{f}_j(\mathbf{x}, t)$ is the post-collision distribution function. It should be noted that the GLBM (\ref{eq:2-1}) has the form of popular MRT model with collision matrix $\mathbf{\Lambda}$ \cite{dHumieres1992}, and the commonly used LB models for the NSEs and nonlinear convection-diffusion equations (NCDEs) can be viewed as its special cases (see section 5 for details). Here we would also like to point out that in almost all of the MRT models, the collision process in Eq. (\ref{eq:2-1}) is carried out in the moment space rather than the velocity (distribution function) space, and the analysis method (e.g., the Chapman-Enskog analysis) is usually more complicated and often depends on the specified lattice structure or discrete velocity set \cite{dHumieres1992,Guo2013,Chai2012,Yoshida2010,Chai2016a}. In this work, we aim to provide a unified framework for the modeling and theoretical analysis of GLBM (\ref{eq:2-1}) in the velocity space.

In the GLBM, besides the evolution equation (\ref{eq:2-1}), some other key elements including the equilibrium distribution function $f_{j}^{eq}$, distribution function of the source or forcing term $F_j$, collision matrix $\mathbf{\Lambda}$ and \textit{auxiliary} source distribution function $G_j$, must also be given properly.
In addition, in the implementation of the GLBM, there are two popular schemes that can be used to discrete the last term on the right hand side of Eq. (\ref{eq:2-1}), i.e., $\Delta t\bar{D}_j F_j(\mathbf{x}, t)/2$. Actually, if $\gamma=0$, the first-order explicit difference scheme $\partial_{t} F_j(\mathbf{x}, t)=[F_j(\mathbf{x}, t+\Delta t)-F_j(\mathbf{x}, t)]/\Delta t$ is adopted for NCDEs \cite{Shi2009,Chai2016a}. While for the case of $\gamma=1$, we can use the first-order implicit difference scheme $(\partial_{t}+\mathbf{c}_{j}\cdot\nabla) F_j(\mathbf{x}, t)=[F_j(\mathbf{x}+\mathbf{c}_{j}\Delta t, t+\Delta t)-F_j(\mathbf{x}, t)]/\Delta t$, and then the linear transform considered in Refs. \cite{He1998, Du2006,Chai2016a} is applied to avoid the implicitness.

\section{The analysis approaches for GLBM: Chapman-Enskog analysis, Maxwell iteration, direct Taylor expansion and recurrence equations}
\label{printlayout}

In the literature, there are mainly four basic analysis approaches for LB models, i.e.,
\begin{enumerate}[(1)]
\item Chapman-Enskog (CE) analysis.
\item Maxwell iteration (MI) method.
\item Direct Taylor expansion (DTE) method.
\item Recurrence equations (RE) method.
\end{enumerate}

In above four approaches, the CE analysis is the most widely used method in the LB community, and combines the physical principle with mathematical technique (Taylor expansion). Different from the CE analysis, however, in the MI, DTE and RE methods, only the Taylor expansion is adopted. In what follows, we will present an introduction to these four analysis methods, and also conduct a comparison among them.

\subsection{The Chapman-Enskog analysis}

To derive the macroscopic equation from Eq. (\ref{eq:2-1}), one can consider the popular CE analysis in which the distribution functions $f_j$, $F_j$, $G_j$, the time and space derivatives can be expanded as \cite{Guo2013,Chai2016a}
\begin{subequations}\label{eq:3-1}
\begin{equation}
f_j=f_j^{(0)}+\epsilon
f_j^{(1)}+\epsilon^2 f_j^{(2)},
\end{equation}
\begin{equation}
G_j=\epsilon
G_j^{(1)}+\epsilon^2 G_j^{(2)}, F_j=\epsilon F_j^{(1)}+\epsilon^2 F_j^{(2)},
\end{equation}
\begin{equation}
\partial_t=\epsilon\partial_{t_1}+\epsilon^2\partial_{t_2},\nabla=\epsilon\nabla_1,
\end{equation}
\end{subequations}
where $\epsilon$ is a small expansion parameter.

Applying the Taylor expansion to Eq. (\ref{eq:2-1}), we have
\begin{equation}\label{eq:3-2}
D_j f_j + \frac{\Delta t}{2}D_j^2 f_j + \cdots = -\frac{1}{\Delta t}\mathbf{\Lambda}_{jk} f_k^{ne}+G_j+F_j+\frac{\Delta t}{2}\bar{D}_j F_j,
\end{equation}
where $D_j=\partial_t+\mathbf{c}_j\cdot\nabla$.
Denoting $D_{1j}=\partial_{t_1}+\mathbf{c}_j\cdot\nabla_1, \bar{D}_{1j}=\partial_{t_1}+\gamma\mathbf{c}_j\cdot\nabla_1$, and
substituting Eq. (\ref{eq:3-1}) into Eq. (\ref{eq:3-2}) yields the following equations at different orders of $\epsilon$,
\begin{subequations}\label{eq:3-3}
\begin{equation}
\epsilon^{0}:\ \ \mathbf{\Lambda}_{jk}\big(f_k^{(0)}-f_k^{eq}\big)=0,
\end{equation}
\begin{equation}
\epsilon^{1}:\ \ D_{1j} f_j^{(0)} = - \frac{1}{\Delta t}\mathbf{\Lambda}_{jk}f_k^{(1)}+G_j^{(1)} +F_j^{(1)},
\end{equation}
\begin{equation}
\epsilon^{2}:\ \ \partial_{t_2} f_j^{(0)} + D_{1j} f_j^{(1)} + \frac{\Delta t}{2} D_{1j}^2 f_j^{(0)} =
-\frac{1}{\Delta t}\mathbf{\Lambda}_{jk} f_k^{(2)}+G_j^{(2)} +F_j^{(2)}+\frac{\Delta t}{2} \bar{D}_{1j} F_j^{(1)}.
\end{equation}
\end{subequations}
Based on the fact that $\mathbf{\Lambda}$ is invertible, Eq. (\ref{eq:3-3}a) gives
\begin{equation}\label{eq:3-4}
f_j^{(0)}=f_j^{eq}, \ j=0,1,\cdots,q-1.
\end{equation}

With the help of Eq. (\ref{eq:3-3}b), we can rewrite Eq. (\ref{eq:3-3}c) as
\begin{equation}\label{eq:3-5}
\partial_{t_2} f_j^{(0)} + D_{1j} \big(\delta_{jk}-\frac{\mathbf{\Lambda}_{jk}}{2}\big)f_k^{(1)}+ \frac{\Delta t}{2}D_{1j} \big(G_j^{(1)}+F_j^{(1)}\big)=-\frac{1}{\Delta t}\mathbf{\Lambda}_{jk} f_k^{(2)}+G_j^{(2)} +F_j^{(2)}+\frac{\Delta t}{2} \bar{D}_{1j} F_j^{(1)}.
\end{equation}
In addition, multiplying $\epsilon$ on both sides of Eq. (\ref{eq:3-3}b) and including some terms at the order of $\epsilon^{2}$, we have
\begin{equation}\label{eq:3-3-2}
D_{j} f_j^{eq} = - \frac{1}{\Delta t}\mathbf{\Lambda}_{jk}f_k^{ne}+G_j +F_j+O(\epsilon^{2}+\epsilon^{2}/\Delta t),
\end{equation}
Through a combination of Eqs. (\ref{eq:3-3}b) and (\ref{eq:3-5}), i.e., $\epsilon^{1}\times$Eq. (\ref{eq:3-3}b)+$\epsilon^{2}\times$Eq. (\ref{eq:3-5}), we can also obtain
\begin{equation}\label{eq:3-6}
D_{j} f_j^{eq} + D_{j} \big(\delta_{jk}-\frac{\mathbf{\Lambda}_{jk}}{2}\big) f_k^{ne}+ \frac{\Delta t}{2}D_{j}\big(G_j+F_j\big)=-\frac{1}{\Delta t}\mathbf{\Lambda}_{jk} f_k^{ne}+G_j +F_j+\frac{\Delta t}{2} \bar{D}_{j} F_j+O(\epsilon^{3}+\epsilon^{3}/\Delta t).
\end{equation}

\subsection{The Maxwell iteration method}

To simplify the following discussion on the MI method, we first introduce the following notations,
\begin{subequations}\label{eq:3-7}
\begin{equation}
\mathbf{f}=(f_0,f_1,\cdots,f_{q-1})^T,\ \ \mathbf{f}^{eq}=(f_0^{eq},f_1^{eq},\cdots,f_{q-1}^{eq})^T,\ \ \mathbf{f}^{ne}=\mathbf{f}-\mathbf{f}^{eq}
\end{equation}
\begin{equation}
\mathbf{G}=(G_0,G_1,\cdots,G_{q-1})^T,\ \ \mathbf{F}=(F_0,F_1,\cdots,F_{q-1})^T,
\end{equation}
\begin{equation}
\bar{\mathbf{D}}=\textbf{diag}(\bar{D}_0,\bar{D}_1,\cdots,\bar{D}_{q-1}),\ \ \mathbb{D}=\textbf{diag}(D_0,D_1,\cdots,D_{q-1}),
\end{equation}
\begin{equation}
\mathbf{L}=\mathbf{L}(\Delta t \mathbb{D})=\sum_{k\geq1} \frac{(\Delta t \mathbb{D})^k}{k!},
\end{equation}
\end{subequations}
then with the Taylor expansion, Eq. (\ref{eq:2-1}) can be written as an operator equation,
\begin{equation}\label{eq:3-8}
\mathbf{Lf}=-\mathbf{\Lambda} \mathbf{f}^{ne} + \Delta t \tilde{\mathbf{F}}=-\mathbf{\Lambda} (\mathbf{f}-\mathbf{f}^{eq}) + \Delta t \tilde{\mathbf{F}}.
\end{equation}
where $\tilde{\mathbf{F}}=\mathbf{G}+\mathbf{F}+(\Delta t/2)\bar{\mathbf{D}} \mathbf{F}$.

From Eq. (\ref{eq:3-8}), one can obtian
\begin{equation}\label{eq:3-9}
\mathbf{f}=\mathbf{f}^{eq}+\Delta t \mathbf{\Lambda}^{-1} \tilde{\mathbf{F}} -\mathbf{\Lambda}^{-1} \mathbf{Lf}.
\end{equation}
Denote
\begin{equation}\label{eq:3-10}
\bar{\mathbf{L}}=\mathbf{\Lambda}^{-1} \mathbf{L},\ \ \bar{\mathbf{G}}=\mathbf{\Lambda}^{-1} \tilde{\mathbf{F}},\ \ \bar{\mathbf{f}}^{eq}=\mathbf{f}^{eq}+\Delta t \mathbf{\Lambda}^{-1} \tilde{\mathbf{F}}=\mathbf{f}^{eq}+\Delta t \bar{\mathbf{G}},
\end{equation}
we can write Eq. (\ref{eq:3-9}) in a simple form,
\begin{equation}\label{eq:3-11}
\mathbf{f}=\bar{\mathbf{f}}^{eq} - \bar{\mathbf{L}}\mathbf{f}.
\end{equation}
In the MI method \cite{Ikenberry1956,Yong2016,Zhao2017,Zhang2019} where the distribution function $\mathbf{f}$ is substituted into the right hand side of Eq. (\ref{eq:3-11}), one can obtain the following formula from Eq. (\ref{eq:3-11}),
\begin{equation}\label{eq:3-15}
\mathbf{f} = \sum_{i=0}^{+\infty} (-\bar{\mathbf{L}})^i \bar{\mathbf{f}}^{eq}= \sum_{i=0}^{k} (-\bar{\mathbf{L}})^i \bar{\mathbf{f}}^{eq}+O(\Delta t^{k+1})=\sum_{i=0}^{k} (-\bar{\mathbf{L}})^i \mathbf{f}^{eq}+\Delta t\sum_{i=0}^{k-1}(-\bar{\mathbf{L}})^i \bar{\mathbf{G}}+O(\Delta t^{k+1}),
\end{equation}
which also leads to
\begin{subequations}\label{eq:3-17}
\begin{equation}
\mathbf{f}=(\mathbf{I}-\bar{\mathbf{L}})\mathbf{f}^{eq}+\Delta t \bar{\mathbf{G}}=(\mathbf{I}-\bar{\mathbf{L}})\mathbf{f}^{eq}+\Delta t \mathbf{\Lambda}^{-1}(\mathbf{G}+\mathbf{F})+O(\Delta t^2),
\end{equation}
\begin{equation}
\mathbf{f}=(\mathbf{I}-\bar{\mathbf{L}}+\bar{\mathbf{L}}^2)\mathbf{f}^{eq}+\Delta t (\mathbf{I}-\bar{\mathbf{L}})\mathbf{\Lambda}^{-1}(\mathbf{G}+\mathbf{F})+\frac{\Delta t^{2}}{2}\mathbf{\Lambda}^{-1}\bar{\mathbf{D}}\mathbf{F}+O(\Delta t^3),
\end{equation}
\end{subequations}
where $\bar{\mathbf{L}}=O(\Delta t)$ has been applied.

In addition, from Eq. (\ref{eq:3-7}) one can also derive
\begin{subequations}\label{eq:3-18}
\begin{eqnarray}
\bar{\mathbf{L}}=\mathbf{\Lambda}^{-1}\mathbf{L} & = & \mathbf{\Lambda}^{-1}\big(\Delta t \mathbb{D}+\frac{\Delta t^2}{2!}\mathbb{D}^2+\frac{\Delta t^3}{3!}\mathbb{D}^3+\cdots\big)\nonumber\\
& = & \mathbf{\Lambda}^{-1}\big(\Delta t \mathbb{D}+\frac{\Delta t^2}{2!}\mathbb{D}^2\big)+O(\Delta t^3)\nonumber\\
& = & \Delta t\mathbf{\Lambda}^{-1} \mathbb{D}+O(\Delta t^2),
\end{eqnarray}
\begin{equation}
\bar{\mathbf{L}}^2=\Delta t^2(\mathbf{\Lambda}^{-1}\mathbb{D})^2+O(\Delta t^3).
\end{equation}
\end{subequations}
Substituting Eq. (\ref{eq:3-18}) into Eq. (\ref{eq:3-17}) yields
\begin{subequations}\label{eq:3-20}
\begin{equation}
\mathbf{f}^{ne}=\mathbf{f}-\mathbf{f}^{eq}=-\Delta t \mathbf{\Lambda}^{-1}\mathbb{D}\mathbf{f}^{eq}+\Delta t \mathbf{\Lambda}^{-1}(\mathbf{G}+\mathbf{F})+O(\Delta t^2),
\end{equation}
\begin{equation}
\mathbf{f}^{ne}=\mathbf{f}-\mathbf{f}^{eq}=-\Delta t \mathbf{\Lambda}^{-1}\mathbb{D}\mathbf{f}^{eq}+\Delta t \bar{\mathbf{G}}+\Delta t^2\mathbf{\Lambda}^{-1}\mathbb{D}\big[\big(\mathbf{I}-\frac{\mathbf{\Lambda}}{2}\big)\Lambda^{-1}\mathbb{D}\mathbf{f}^{eq}-\Lambda^{-1}(\mathbf{G}+\mathbf{F})\big]+O(\Delta t^3).
\end{equation}
\end{subequations}
Based on Eq. (\ref{eq:3-20}) and using $\bar{\mathbf{G}}=\mathbf{\Lambda}^{-1}\tilde{\mathbf{F}}=\mathbf{\Lambda}^{-1}(\mathbf{G}+\mathbf{F})+O(\Delta t)$, we have
\begin{subequations}\label{eq:3-21}
\begin{equation}
\mathbf{f}^{ne}=O(\Delta t),
\end{equation}
\begin{equation}
\mathbb{D}\mathbf{f}^{eq}=-\frac{\mathbf{\Lambda}}{\Delta t}\mathbf{f}^{ne}+ \mathbf{G}+\mathbf{F} +O(\Delta t),
\end{equation}
\begin{equation}
\mathbb{D}\mathbf{f}^{eq}-\Delta t \mathbb{D}\big[\big(\mathbf{I}-\frac{\mathbf{\Lambda}}{2}\big)\mathbf{\Lambda}^{-1}\mathbb{D}\mathbf{f}^{eq}-\mathbf{\Lambda}^{-1}(\mathbf{G}+\mathbf{F})\big]=-\frac{\mathbf{\Lambda}}{\Delta t}\mathbf{f}^{ne}+ \mathbf{G}+\mathbf{F} +\frac{\Delta t}{2}\bar{\mathbf{D}}\mathbf{F}+O(\Delta t^2).
\end{equation}
\end{subequations}
Inserting Eq. (\ref{eq:3-21}b) into Eq. (\ref{eq:3-21}c) gives rise to the following equation,
\begin{equation}\label{eq:3-22}
\mathbb{D}\mathbf{f}^{eq}+ \mathbb{D}\big(\mathbf{I}-\frac{\mathbf{\Lambda}}{2}\big)\mathbf{f}^{ne}+\frac{\Delta t}{2}\mathbb{D}(\mathbf{G}+\mathbf{F})=-\frac{\mathbf{\Lambda}}{\Delta t}\mathbf{f}^{ne}+\mathbf{G}+\mathbf{F} +\frac{\Delta t}{2}\bar{\mathbf{D}}\mathbf{F}+O(\Delta t^2).
\end{equation}
We note that Eqs. (\ref{eq:3-21}b) and (\ref{eq:3-22}) are the same as Eqs. (\ref{eq:3-3-2}) and (\ref{eq:3-6}) in the CE analysis except the truncation errors neglected. Actually, when $\epsilon=O(\Delta t)$, the truncation errors of the CE analysis and MI method are of the same order in $\Delta t$.

\subsection{The direct Taylor expansion method}

From above discussion, it can be seen clearly that Eqs. (\ref{eq:3-22}) and (\ref{eq:3-6}) have the similar form. In addition, one can also find that the aim of the MI method is to give an approximate expression of distribution function in terms of $\bar{\mathbf{f}}^{eq}$ or $\mathbf{f}^{eq}$ [see Eq. (\ref{eq:3-15})], which usually makes the MI method a little more complicated than the CE expansion. However, above analysis [see Eqs. (\ref{eq:3-21}) and (\ref{eq:3-22})] based on the MI method also implies that it may be possible to get the same results by directly using the Taylor expansion. In the following, we present this simple approach called direct Taylor expansion method, which is similar to those adopted in the previous works \cite{Kaehler2013,Holdych2004,Wagner2006}.

The basic idea of the DTE method is to give a series of equations on $\mathbf{f}^{eq}$ and $\mathbf{f}^{ne}$ with different orders of truncation errors from Eq. (\ref{eq:2-1}), which is different from that in the MI method.
If we apply the Taylor expansion to Eq. (\ref{eq:2-1}), one can get
\begin{equation}\label{eq:3-23}
\sum_{l=1}^{N}\frac{\Delta t^l}{l!}D_j^l f_j +O(\Delta t^{N+1})=-\mathbf{\Lambda}_{jk} f_k^{ne}+\Delta t \tilde{F}_j,
\end{equation}
Based on the relation $f_j=f_j^{eq}+f_j^{ne}$ and Eq. (\ref{eq:3-23}), the following equations are obtained,
\begin{subequations}\label{eq:3-24}
\begin{equation}
f_j^{ne}=O(\Delta t),
\end{equation}
\begin{equation}
\sum_{l=1}^{N-1}\frac{\Delta t^l}{l!}D_j^l (f_j^{eq}+f_j^{ne}) + \frac{\Delta t^N}{N!}D_j^N f_j^{eq}=-\mathbf{\Lambda}_{jk} f_k^{ne}+\Delta t \tilde{F}_j+O(\Delta t^{N+1}).
\end{equation}
\end{subequations}
Then from Eq. (\ref{eq:3-24}), one can derive the equations at different orders of $\Delta t$,
\begin{subequations}\label{eq:3-25}
\begin{equation}
\Delta t D_j f_j^{eq}=-\mathbf{\Lambda}_{jk} f_k^{ne}+\Delta t (G_j+F_j)+O(\Delta t^2),
\end{equation}
\begin{equation}
\Delta t D_j (f_j^{eq}+f_j^{ne})+\frac{\Delta t^2}{2}D_j^2 f_j^{eq}=-\mathbf{\Lambda}_{jk} f_k^{ne}+\Delta t \big(G_j+F_j+\frac{\Delta t}{2}\bar{D}_jF_j\big)+O(\Delta t^3),
\end{equation}
\end{subequations}
or equivalently,
\begin{subequations}\label{eq:3-26}
\begin{equation}
D_j f_j^{eq}=-\frac{\mathbf{\Lambda}_{jk}}{\Delta t } f_k^{ne}+G_j+F_j+O(\Delta t),
\end{equation}
\begin{equation}
D_j (f_j^{eq}+f_j^{ne})+\frac{\Delta t}{2}D_j^2 f_j^{eq}=-\frac{\mathbf{\Lambda}_{jk}}{\Delta t } f_k^{ne}+ G_j+F_j+\frac{\Delta t}{2}\bar{D}_jF_j+O(\Delta t^2).
\end{equation}
\end{subequations}
According to Eq. (\ref{eq:3-25}a), we have
\begin{subequations}\label{eq:3-27}
\begin{equation}
D_j f_j^{ne}=-\Delta t D_j \mathbf{\Lambda}_{jk}^{-1}\big[D_k f_k^{eq}-(G_k+F_k)\big]+O(\Delta t^2),
\end{equation}
\begin{equation}
\frac{\Delta t}{2}D_j^2 f_j^{eq}=-\frac{1}{2} D_j \mathbf{\Lambda}_{jk} f_k^{ne}+\frac{\Delta t}{2}D_j(G_j+F_j)+O(\Delta t^2).
\end{equation}
\end{subequations}
Substituting Eq. (\ref{eq:3-27}) into Eq. (\ref{eq:3-26}b), one can obtain the following two equivalent equations,
\begin{equation}\label{eq:3-28}
D_j f_j^{eq}-\Delta t D_j\big[\big(\delta_{jl}-\frac{\mathbf{\Lambda}_{jl}}{2}\big)\mathbf{\Lambda}_{lk}^{-1}D_k f_k^{eq}-\mathbf{\Lambda}_{jk}^{-1}(G_k+F_k)\big]=-\frac{\mathbf{\Lambda}_{jk}}{\Delta t} f_k^{ne}+G_j+F_j+\frac{\Delta t}{2}\bar{D}_jF_j+O(\Delta t^2),
\end{equation}
\begin{equation}\label{eq:3-29}
D_j f_j^{eq}+D_j\big(\delta_{jk}-\frac{\mathbf{\Lambda}_{jk}}{2}\big)f_k^{ne}+\frac{\Delta t}{2}D_j(G_j+F_j)\nonumber\\=-\frac{\mathbf{\Lambda}_{jk}}{\Delta t} f_k^{ne}+G_j+F_j+\frac{\Delta t}{2}\bar{D}_jF_j+O(\Delta t^2).
\end{equation}
It is worth noting that Eqs. (\ref{eq:3-26}a) and (\ref{eq:3-29}) are the same as Eqs. (\ref{eq:3-21}b) and (\ref{eq:3-22}) which means that mathematically, the DTE method is equivalent to MI method, while the DTE method seems more straightforward and much simpler. In addition, we would also like to point out that different from the DTE method used in Ref. \cite{Kaehler2013,Wagner2006}, we do not make any assumptions in above analysis.

\subsection{The recurrence equations method}

Like MI method, the RE method is also an approach based on Taylor expansion. In the RE method \cite{dHumieres2009,Ginzburg2015}, a kind of difference equations about $f_j^{eq}$ and $f_j^{ne}$ named recurrence equations is first constructed from the evolution equation of LB model, then one can apply Taylor expansion to these recurrence equations to obtain the serious equations at different orders of truncation errors.

In order to derive REs, we first rewrite Eq. (\ref{eq:2-1}) as
\begin{subequations} \label{eq:3-30}
\begin{equation}
f_j(\mathbf{x}+\mathbf{c}_j \Delta t,t)=\tilde{f}_j(\mathbf{x}, t-\Delta t),
\end{equation}
\begin{equation}
\tilde{f}_j(\mathbf{x}-\mathbf{c}_j \Delta t,t)=f_j(\mathbf{x}, t+\Delta t).
\end{equation}
\end{subequations}
Let
\begin{equation}\label{eq:3-31}
\mathbf{g}=-\mathbf{\Lambda} \mathbf{f}^{ne},\ \ \mathbf{\bar{g}}=\mathbf{g}+\Delta t\tilde{\mathbf{F}}=-\mathbf{\Lambda} \bar{\mathbf{f}}^{ne}=-\mathbf{\Lambda}( \mathbf{f}-\bar{\mathbf{f}}^{eq}),\ \ \mathbf{\Lambda}^{-1}=\mathbf{I}/2+\bar{\mathbf{\Lambda}},
\end{equation}
then
\begin{equation}\label{eq:3-32}
\mathbf{f}^{ne}=-\mathbf{\Lambda}^{-1} \mathbf{g},\ \ \bar{\mathbf{f}}^{ne}=-\mathbf{\Lambda}^{-1}\mathbf{\bar{g}},\ \ \mathbf{I}-\mathbf{\Lambda}^{-1}=\mathbf{I}/2-\bar{\mathbf{\Lambda}},
\end{equation}
which can be used to derive the following equations,
\begin{subequations} \label{eq:3-33}
\begin{equation}
f_j=\bar{f}_j^{eq}+\bar{f}_j^{ne}=\bar{f}_j^{eq}-(\mathbf{\Lambda}^{-1})_{jk}\bar{g}_k=\bar{f}_j^{eq}-\big(\frac{\delta_{jk}}{2}+\bar{\mathbf{\Lambda}}_{jk}\big)\bar{g}_k,
\end{equation}
\begin{equation}
\tilde{f_j}=\bar{f}_j^{eq}+(\delta_{jk}-\mathbf{\Lambda}_{jk})\bar{f}_k^{ne}=\bar{f}_j^{eq}+\big(\frac{\delta_{jk}}{2}-\bar{\mathbf{\Lambda}}_{jk}\big)\bar{g}_k.
\end{equation}
\end{subequations}
Substituting Eq. (\ref{eq:3-33}) into Eq. (\ref{eq:3-30}), one can obtain
\begin{subequations} \label{eq:3-34}
\begin{equation}
 \big[\bar{f}_j^{eq}-\big(\frac{\delta_{jk}}{2}+\bar{\mathbf{\Lambda}}_{jk}\big)\bar{g}_k\big](\mathbf{x}+\mathbf{c}_j \Delta t,t)
=\big[\bar{f}_j^{eq}+\big(\frac{\delta_{jk}}{2}-\bar{\mathbf{\Lambda}}_{jk}\big)\bar{g}_k\big](\mathbf{x},t-\Delta t),
\end{equation}
\begin{equation}
 \big[\bar{f}_j^{eq}+\big(\frac{\delta_{jk}}{2}-\bar{\mathbf{\Lambda}}_{jk}\big)\bar{g}_k\big](\mathbf{x}-\mathbf{c}_j \Delta t,t)
=\big[\bar{f}_j^{eq}-\big(\frac{\delta_{jk}}{2}+\bar{\mathbf{\Lambda}}_{jk}\big)\bar{g}_k\big](\mathbf{x},t+\Delta t).
\end{equation}
\end{subequations}
The sum and the difference of Eqs. (\ref{eq:3-34}a) and (\ref{eq:3-34}b) become \cite{dHumieres2009,Ginzburg2015}
\begin{subequations} \label{eq:3-35}
\begin{equation}
\big[(\Delta_j^2-\Delta_t^2)(\bar{f}_j^{eq}-\bar{\mathbf{\Lambda}}_{jk}\bar{g}_k)-(\bar{\Delta}_j-\bar{\Delta}_t)\bar{g}_j\big](\mathbf{x},t)
=0,
\end{equation}
\begin{equation}
\big[(\bar{\Delta}_j+\bar{\Delta}_t)(\bar{f}_j^{eq}-\bar{\mathbf{\Lambda}}_{jk}\bar{g}_k)-\frac{1}{4}(\Delta_j^2+\Delta_t^2)\bar{g}_j\big](\mathbf{x},t)
=\bar{g}_j(\mathbf{x},t),
\end{equation}
\end{subequations}
where the following central difference schemes are adopted for the time and space derivatives of the variable $\phi=\{\bar{f}_j^{eq},\bar{g}_j\}$,
\begin{subequations} \label{eq:3-36}
\begin{equation}
\bar{\Delta}_t \phi(\mathbf{x},t)=[\phi(\mathbf{x},t+\Delta t)-\phi(\mathbf{x},t-\Delta t)]/2,
\end{equation}
\begin{equation}
\Delta_t^2 \phi(\mathbf{x},t)=\phi(\mathbf{x},t+\Delta t)-2\phi(\mathbf{x},t)+\phi(\mathbf{x},t-\Delta t),
\end{equation}
\begin{equation}
\bar{\Delta}_j \phi(\mathbf{x},t)=[\phi(\mathbf{x}+\mathbf{c}_j \Delta t,t)-\phi(\mathbf{x}-\mathbf{c}_j \Delta t,t)]/2,
\end{equation}
\begin{equation}
\Delta_j^2 \phi(\mathbf{x},t)=\phi(\mathbf{x}+\mathbf{c}_j \Delta t,t)-2\phi(\mathbf{x},t)+\phi(\mathbf{x}-\mathbf{c}_j \Delta t,t).
\end{equation}
\end{subequations}
Here It should be noted that unlike the previous works limited to cases with the constant coefficients \cite{dHumieres2009,Ginzburg2015}, the results of the present RE method can be used for a more general case with variable coefficients.

In addition, we would like to point out that Eqs. (\ref{eq:3-34}) and (\ref{eq:3-35}) are the so-called REs obtained from Eq. (\ref{eq:2-1}). Through using Taylor expansion to the REs, we can also determine a series of equations on $f_j^{eq}$ and $g_j$ (or $f_j^{ne}$) with different orders of time step. Actually, applying the Taylor expansion to Eq. (\ref{eq:3-35}b) gives
\begin{equation}\label{eq:3-38a}
D_j \big(\bar{f}_j^{eq}-\bar{\mathbf{\Lambda}}_{jk}\bar{g}_k\big)=\frac{1}{\Delta t}\bar{g}_j+O(\Delta t^2),
\end{equation}
Based on the relation $\bar{g}_j=O(\Delta t)$ [see Eq. (\ref{eq:3-31})], we have
\begin{equation}\label{eq:3-38b}
 D_j \bar{f}_j^{eq}=\frac{1}{\Delta t}\bar{g}_j+O(\Delta t).
\end{equation}
Then we can also rewrite Eqs. (\ref{eq:3-38b}) and (\ref{eq:3-38a}) as
\begin{subequations}\label{eq:3-38}
\begin{equation}
 D_j f_j^{eq}=\frac{1}{\Delta t}g_j+G_j+F_j+O(\Delta t),
\end{equation}
\begin{equation}
D_j \big(f_j^{eq}-\bar{\mathbf{\Lambda}}_{jk}g_k\big)+\frac{\Delta t}{2}D_j \tilde{F}_j=\frac{1}{\Delta t}g_j+\tilde{F}_j+O(\Delta t^2),
\end{equation}
\end{subequations}
where Eqs. (\ref{eq:3-10}), (\ref{eq:3-31}) and (\ref{eq:3-32}) have been used.
Keeping in mind that $\tilde{F}_j=G_j+F_j+(\Delta t/2)\bar{D}_jF_j$, the following equation can be obtained from Eq. (\ref{eq:3-38}b),
\begin{equation}\label{eq:3-39}
 D_j f_j^{eq}-D_j\bar{\mathbf{\Lambda}}_{jk} g_k+\frac{\Delta t}{2}D_j (G_j+F_j)=\frac{1}{\Delta t}g_j+G_j+F_j+\frac{\Delta t}{2}\bar{D}_jF_j+O(\Delta t^2).
\end{equation}
It should be noted that Eqs. (\ref{eq:3-38}a) and (\ref{eq:3-39}) are the same as Eqs. (\ref{eq:3-26}a) and (\ref{eq:3-29}), which also means that the MI, DTE and RE methods are equivalent to each other.

\section{Derivation of the macroscopic equations}

In this section, we will adopt above analysis methods to derive the macroscopic equations including the nonlinear anisotropic convection-diffusion equation (NACDE) and classical NSEs, and mainly focus on the CE analysis and some of other methods. This is because from the discussion in the previous section, we can see that the CE method has incorporated some \emph{physical} principles, and compared to some other \emph{mathematical} methods based Taylor expansion, it also presents some different results at the higher order equations. On the other hand, it is clear that the MI, DTE and RE methods are equivalent to each other since they can give the same equations at the first and second order of time step, thus for brevity, only some of them are considered.

\subsection{Derivation of NACDE}

The $d$-dimensional NACDE with a source term can be expressed as \cite{Chai2016a}
\begin{equation}\label{eq:4-1}
\partial_t \phi+\nabla\cdot\mathbf{B}= \nabla\cdot[\mathbf{K}\cdot(\nabla\cdot \mathbf{D})]+S,
\end{equation}
where $\phi$ is a unknown scalar function of position $\mathbf{x}$ and time $t$, $S$ is a scalar source term. $\mathbf{B}=(\mathbf{B}_\alpha)$ is a vector function, $\mathbf{K}=(\mathbf{K}_{\alpha\beta})$ and $\mathbf{D}=(\mathbf{D}_{\alpha\beta})$ are symmetric tensors (matrices), and they can be functions of $\phi, \mathbf{x}$, and $t$. We note that Eq. (\ref{eq:4-1}) can be considered as a general form of CDEs, and many different kinds of CDEs considered in some previous works \cite{Ginzburg2005a,Shi2009} are its special cases. When $\mathbf{K}_{\alpha\beta}=\kappa \delta_{\alpha\beta}$ and $\mathbf{D}_{\alpha\beta}=h \delta_{\alpha\beta}$ with $\kappa$ and $h$ being two scalar functions, Eq. (\ref{eq:4-1}) becomes an isotropic CDE.

To recover Eq. (\ref{eq:4-1}) from the GLBM (\ref{eq:2-1}), we only need to give some appropriate constraints on the collision matrix $\mathbf{\Lambda}$, the moments of $f_j^{eq}(\mathbf{x},t)$, $G_j(\mathbf{x},t)$ and $F_j(\mathbf{x},t)$. Based on these constrains, one can determine the expressions of $\mathbf{\Lambda}$ and these distribution functions. Actually, to derive Eq. (\ref{eq:4-1}), only the zeroth- to second-order moments of $f_j^{eq}$, the zeroth- and first-order moments of $G_j$ and $F_j$, and the constraints on $\mathbf{\Lambda}$ corresponding to these moments are needed. In the GLBM, the unknown conserved scalar $\phi$ can be calculated by $\phi=\sum_j f_j$, the distribution functions $f_j$, $f_j^{eq}$, $G_j$ and $F_j$ should satisfy the following relations,

\begin{subequations}\label{eq:4-2}
\begin{equation}
\sum_j f_j=\sum_j f_j^{eq}=\phi,\ \ \sum_j \mathbf{c}_j f_j^{eq}=\mathbf{B},\ \ \sum_j \mathbf{c}_j \mathbf{c}_j
f_j^{eq}=\beta c_s^2 \mathbf{D}+\mathbf{C},
\end{equation}
\begin{equation}
\sum_j G_j=0,\ \ \sum_j \mathbf{c}_j G_j=\mathbf{M}_{1G},
\end{equation}
\begin{equation}
\sum_j F_j=S,\ \ \sum_j \mathbf{c}_j F_j=\mathbf{M}_{1F},
\end{equation}
\end{subequations}
where $c_s$ is the lattice sound speed related to lattice speed $c=\Delta x/\Delta t$ with $\Delta x$ being the lattice spacing. $\beta$ is a parameter that can be used to adjust the relaxation matrix [see following Eq. (\ref{eq:4-10})], $\mathbf{C}$ is an auxiliary moment to be determined later, $\mathbf{M}_{1G}$ and $\mathbf{M}_{1F}$ are the first-order moments of $G_j$ and $F_j$.

In addition, for the collision matrix $\mathbf{\Lambda}$, the following requirements are needed,
\begin{eqnarray}\label{eq:4-3}
\sum_j \mathbf{e}_{j}\mathbf{\Lambda}_{jk} = s_0 \mathbf{e}_{k},\ \ \sum_j \mathbf{c}_{j}\mathbf{\Lambda}_{jk} = \mathbf{S}_1 \mathbf{c}_k,\ \ \forall k,
\end{eqnarray}
where $\mathbf{e}=(1,1,\cdots,1)\in R^q$, $\mathbf{S}_1$ is an invertible $d\times d$ relaxation matrix corresponding to the diffusion matrix $\mathbf{K}$. Based on Eqs. (\ref{eq:4-2}a) and (\ref{eq:4-3}), one can find that the relaxation parameter $s_{0}$ does not appear in the recovered macroscopic equation, and its value can be chosen arbitrarily.

\subsubsection{Derivation of the NACDE through CE analysis}

In this part, we will present some details on how to derive NACDE (\ref{eq:4-1}) from GLBM (\ref{eq:2-1}).
In the CE analysis, we can first express the source term as $S=\epsilon S^{(1)}+\epsilon^2 S^{(2)}$, then from Eqs. (\ref{eq:3-1}), (\ref{eq:3-4}) and (\ref{eq:4-2}) we have
\begin{subequations}\label{eq:4-4}
\begin{equation}
\sum_j f_j^{(k)}=0,
\end{equation}
\begin{equation}
\sum_j G_j^{(k)}=0,\ \ \sum_j \mathbf{c}_j G_j^{(k)}=\mathbf{M}_{1G}^{(k)},
\end{equation}
\begin{equation}
\sum_j F_j^{(k)}=S^{(k)},\ \ \sum_j \mathbf{c}_j F_j^{(k)}=\mathbf{M}_{1F}^{(k)},
\end{equation}
\end{subequations}
where $k\geq 1$. Summing Eqs. (\ref{eq:3-3}b) and (\ref{eq:3-5}) over $j$ and using Eqs. (\ref{eq:4-2}), (\ref{eq:4-3}) and (\ref{eq:4-4}), one can obtain
\begin{subequations}\label{eq:4-5}
\begin{equation}
\partial_{t_1} \phi +  \nabla_1\cdot\mathbf{B}= S^{(1)},
\end{equation}
\begin{equation}
\partial_{t_2} \phi +\nabla_1\cdot \big[(\mathbf{I}-\mathbf{S}_1/2)\sum_k \mathbf{c}_k f_k^{(1)}\big]= S^{(2)}+\frac{\Delta t}{2}\nabla_1\cdot\big[(\gamma-1)\mathbf{M}_{1F}^{(1)}-\mathbf{M}_{1G}^{(1)}\big].
\end{equation}
\end{subequations}
With the aid of Eqs. (\ref{eq:3-3}b), (\ref{eq:4-2}), (\ref{eq:4-3}) and (\ref{eq:4-4}), we get
\begin{eqnarray}\label{eq:4-6}
\sum_k \mathbf{c}_k f_k^{(1)}& = &-\Delta t \sum_k \mathbf{c}_k\mathbf{\Lambda}_{kl}^{-1}\big(D_{1l}f_l^{(0)}-G_l^{(1)}-F_l^{(1)}\big)\nonumber\\
& = &-\Delta t \mathbf{S}_1^{-1} \sum_l \mathbf{c}_l \big(D_{1l}f_l^{(0)}-G_l^{(1)}-F_l^{(1)} \big)\nonumber\\
& = &-\Delta t \mathbf{S}_1^{-1} \big[\partial_{t_1}\mathbf{B}+\nabla_1\cdot(\beta c_s^2 \mathbf{D}+\mathbf{C})- \mathbf{M}_{1G}^{(1)}- \mathbf{M}_{1F}^{(1)}\big].
\end{eqnarray}
Substituting above equation into Eq. (\ref{eq:4-5}b), we have
\begin{equation}\label{eq:4-7}
\partial_{t_2} \phi = \nabla_1\cdot \big[\Delta t \beta c_s^2 \big(\mathbf{S}_1^{-1}-\mathbf{I}/2\big)\nabla_1\cdot \mathbf{D}\big] +S^{(2)}
+\Delta t\nabla_1\cdot RH,
\end{equation}
where
\begin{equation}\label{eq:4-8}
RH=(\mathbf{S}_1^{-1}-\mathbf{I}/2)(\partial_{t_1}\mathbf{B}+\nabla_1\cdot \mathbf{C})-\big(\mathbf{S}_1^{-1}-\gamma \mathbf{I}/2\big)\mathbf{M}_{1F}^{(1)}-\mathbf{S}_1^{-1}\mathbf{M}_{1G}^{(1)}.
\end{equation}
If we take $RH=0$, Eq. (\ref{eq:4-7}) would reduce to
\begin{equation}\label{eq:4-9}
\partial_{t_2} \phi = \nabla_1\cdot [\mathbf{K}\cdot(\nabla_1\cdot \mathbf{D})] +S^{(2)},
\end{equation}
with
\begin{equation}\label{eq:4-10}
\mathbf{K}=\Delta t \beta c_s^2(\mathbf{S}_1^{-1}-\mathbf{I}/2).
\end{equation}
Combining Eqs. (\ref{eq:4-5}a) with (\ref{eq:4-9}), we obtain the macroscopic NACDE (\ref{eq:4-1}).

Here we would like to point out that in the expression of diffusion matrix $\mathbf{K}$ [see Eq. (\ref{eq:4-10})], the part $(-\Delta t \beta c_s^2\mathbf{I}/2)$ is the numerical diffusion caused by the discrete effect. Actually, when we use the CE analysis to obtain the NACDE from the discrete-velocity Boltzmann equation (see Appendix B for details), the term $(-\Delta t \beta c_s^2\mathbf{I}/2)$ is not included in the matrix $\mathbf{K}$.
In addition, from above CE analysis, we can also see that the functions $\mathbf{C}, \mathbf{M}_{1G}$ and $\mathbf{M}_{1F}$ must be chosen properly to make $RH=0$, and in this case, the NACDE (\ref{eq:4-1}) can be recovered correctly without additional assumptions. In the following, we present some special cases.

\noindent
\textbf{Case 1:} $\mathbf{C}=0,\ \mathbf{M}_{1F}=0$. Under the condition of $RH=0$, we have
\begin{equation}\label{eq:4-11}
RH=\big[(\mathbf{S}_1^{-1}-\mathbf{I}/2)\partial_{t_1}\mathbf{B}-\mathbf{S}_1^{-1}\mathbf{M}_{1G}^{(1)}\big]=0,
\end{equation}
which leads to the following $\mathbf{M}_{1G}$,
\begin{equation}\label{eq:4-12}
\mathbf{M}_{1G}=(\mathbf{I}-\mathbf{S}_1/2)\partial_{t}\mathbf{B}.
\end{equation}

\noindent
\textbf{Case 2:} $\mathbf{B}=\mathbf{B}(\phi),\ \mathbf{M}_{1F}=0$. If $\mathbf{B}$ is a differentiable function of $\phi$, $\mathbf{C}$ and $\mathbf{M}_{1G}$ should satisfy the following equation,
\begin{equation}\label{eq:4-13}
RH=\big[(\mathbf{S}_1^{-1}-\mathbf{I}/2)(\partial_{t_1}\mathbf{B}+\nabla_1\cdot \mathbf{C})-\mathbf{S}_1^{-1}\mathbf{M}_{1G}^{(1)}\big]=0,
\end{equation}
which leads to
\begin{equation}\label{eq:4-13a}
\mathbf{M}_{1G}^{(1)}=(\mathbf{I}-\mathbf{S}_1/2)(\partial_{t_1}\mathbf{B}+\nabla_1\cdot \mathbf{C}).
\end{equation}
Furthermore, if we take $\mathbf{C}$ as
\begin{equation}\label{eq:4-14}
\mathbf{C}_{\alpha\beta}=\int\mathbf{B}_\alpha'\mathbf{B}_\beta'd\phi,
\end{equation}
which satisfies
\begin{equation}\label{eq:4-15}
\mathbf{C}_{\alpha\beta}'=\mathbf{B}_\alpha'\mathbf{B}_\beta',
\end{equation}
then it follows from Eq. (\ref{eq:4-5}a) that
\begin{equation}\label{eq:4-16}
\partial_{t_1}\mathbf{B}+\nabla_1\cdot \mathbf{C}=\mathbf{B}'(\partial_{t_1}\phi+\nabla_1\cdot \mathbf{B})=\mathbf{B}'S^{(1)}.
\end{equation}
Substituting above equation into Eq. (\ref{eq:4-13}), one can determine $\mathbf{M}_{1G}$ by
\begin{equation}\label{eq:4-17}
\mathbf{M}_{1G}=(\mathbf{I}-\mathbf{S}_1/2)\mathbf{B}'S.
\end{equation}

\subsubsection{Derivation of NACDE through MI method}

Based on the similarity between CE analysis and MI method, we can also recover NACDE (\ref{eq:4-1}) through MI method. To do this, we first introduce the following notations,
\begin{subequations}\label{eq:4-18}
\begin{equation}
\mathbf{E}=(\mathbf{c}_0,\mathbf{c}_1,\cdots,\mathbf{c}_{q-1}),
\end{equation}
\begin{equation}
\langle \mathbf{EE} \rangle=(\mathbf{c}_0\mathbf{c}_0,\mathbf{c}_1\mathbf{c}_1,\cdots,\mathbf{c}_{q-1}\mathbf{c}_{q-1}),
\end{equation}
\end{subequations}
then Eqs. (\ref{eq:4-2}) and (\ref{eq:4-3}) can be written in the matrix forms,
\begin{subequations}\label{eq:4-19}
\begin{equation}
\phi=\mathbf{e}\cdot \mathbf{f}=\mathbf{e}\cdot \mathbf{f}^{eq},\ \ \mathbf{E}\mathbf{f}^{eq}=\mathbf{B},\ \langle \mathbf{EE} \rangle \mathbf{f}^{eq}=\beta c_s^2 \mathbf{D}+\mathbf{C},
\end{equation}
\begin{equation}
\mathbf{e}\cdot \mathbf{G}=0,\ \mathbf{E} \mathbf{G}=\mathbf{M}_{1G},\ \ \mathbf{e}\cdot \mathbf{F}=S,\ \ \mathbf{EF}=\mathbf{M}_{1F}=\mathbf{0},
\end{equation}
\begin{equation}
\mathbf{e}\mathbf{\Lambda}=s_0 \mathbf{e},\ \ \mathbf{E}\mathbf{\Lambda}=\mathbf{S}_1\mathbf{E}.
\end{equation}
\end{subequations}
According to above equations, we have
\begin{subequations}\label{eq:4-20}
\begin{equation}
\mathbf{e}\cdot \mathbf{f}^{ne}=0,
\end{equation}
\begin{equation}
\mathbf{e}\cdot \mathbb{D}\mathbf{f}^{eq}=\partial_t \mathbf{e}\cdot \mathbf{f}^{eq}+\nabla \cdot \mathbf{Ef}^{eq}=\partial_t \phi+\nabla\cdot \mathbf{B},
\end{equation}
\begin{equation}
\mathbf{e}\cdot \mathbb{D}(\mathbf{G}+\mathbf{F})=\partial_t \mathbf{e}\cdot (\mathbf{G}+\mathbf{F})+\nabla \cdot \mathbf{E}(\mathbf{G}+\mathbf{F})=\partial_t S+\nabla\cdot \mathbf{M}_{1G},
\end{equation}
\begin{equation}
\mathbf{e}\cdot \bar{\mathbf{D}}\mathbf{F}=\partial_t \mathbf{e}\cdot \mathbf{F}+\gamma\nabla \cdot \mathbf{EF}=\partial_t S,
\end{equation}
\begin{equation}
\mathbf{E}\mathbb{D}\mathbf{f}^{eq}=\partial_t \mathbf{Ef}^{eq}+\nabla\cdot\langle \mathbf{EE} \rangle \mathbf{f}^{eq}=\partial_t\mathbf{B}+\nabla\cdot(\beta c_s^2 \mathbf{D}+\mathbf{C}),
\end{equation}
\begin{eqnarray}
\mathbf{e}\cdot \mathbb{D}(\mathbf{I}-\mathbf{\Lambda}/2)\mathbf{f}^{ne}& = &\partial_t \mathbf{e}\cdot(\mathbf{I}-\mathbf{\Lambda}/2) \mathbf{f}^{ne}+\nabla \cdot \mathbf{E}(\mathbf{I}-\mathbf{\Lambda}/2)\mathbf{f}^{ne}\nonumber\\
& = &\partial_t (1-s_0/2)\mathbf{e}\cdot \mathbf{f}^{ne}+\nabla \cdot (\mathbf{I}-\mathbf{S}_1/2)\mathbf{Ef}^{ne}\nonumber\\
& = &\nabla \cdot (\mathbf{I}-\mathbf{S}_1/2)\mathbf{Ef}^{ne}.
\end{eqnarray}
\end{subequations}
Multiplying $\mathbf{e}$ on both sides of Eqs. (\ref{eq:3-21}b) and (\ref{eq:3-22}), and using Eq. (\ref{eq:4-20}), we have
\begin{subequations}\label{eq:4-21}
\begin{eqnarray}
\mathbf{e}\cdot \mathbb{D}\mathbf{f}^{eq} & = & -\frac{1}{\Delta t}\mathbf{e}\cdot\mathbf{\Lambda f}^{ne}+\mathbf{e}\cdot \mathbf{G}+\mathbf{e}\cdot \mathbf{F}+O(\Delta t)\nonumber\\
& = &-\frac{s_0}{\Delta t}\mathbf{e}\cdot \mathbf{f}^{ne}+S+O(\Delta t),
\end{eqnarray}
\begin{eqnarray}
\mathbf{e}\cdot \mathbb{D}\mathbf{f}^{eq}+\mathbf{e}\cdot \mathbb{D}(\mathbf{I}-\mathbf{\Lambda}/2)\mathbf{f}^{ne}+\frac{\Delta t}{2}\mathbf{e}\cdot \mathbb{D}(\mathbf{G}+\mathbf{F})
& = &-\frac{1}{\Delta t}\mathbf{e}\mathbf{\Lambda}\mathbf{f}^{ne}+\mathbf{e}\cdot \mathbf{G}+\mathbf{e}\cdot \mathbf{F}+\frac{\Delta t}{2}\mathbf{e}\cdot \bar{\mathbf{D}}\mathbf{F}+O(\Delta t^2)\nonumber\\
& = &-\frac{s_0}{\Delta t}\mathbf{e}\cdot \mathbf{f}^{ne}+S+\frac{\Delta t}{2}\mathbf{e}\cdot \bar{\mathbf{D}}\mathbf{F}+O(\Delta t^2).
\end{eqnarray}
\end{subequations}
Based on Eqs. (\ref{eq:4-19}) and (\ref{eq:4-20}), Eq. (\ref{eq:4-21}) can be written as
\begin{subequations}\label{eq:4-22}
\begin{equation}
\partial_t \phi+\nabla\cdot \mathbf{B} = S+O(\Delta t),
\end{equation}
\begin{equation}
\partial_t \phi+\nabla\cdot \mathbf{B} + \nabla \cdot (\mathbf{I}-\mathbf{S}_1/2)\mathbf{Ef}^{ne}+\frac{\Delta t}{2}\nabla\cdot \mathbf{M}_{1G}= S+O(\Delta t^2).
\end{equation}
\end{subequations}
On the other hand, Eq. (\ref{eq:3-21}b) gives
\begin{eqnarray}\label{eq:4-23}
\mathbf{Ef}^{ne}&=&-\Delta t \mathbf{E\Lambda}^{-1}(\mathbb{D}\mathbf{f}^{eq}-\mathbf{G}-\mathbf{F})+O(\Delta t^2)\nonumber\\
&=&-\Delta t \mathbf{S}_1^{-1}\mathbf{E}(\mathbb{D}\mathbf{f}^{eq}-\mathbf{G}-\mathbf{F})+O(\Delta t^2)\nonumber\\
&=&-\Delta t \mathbf{S}_1^{-1}\big[\partial_t \mathbf{B}+\nabla\cdot(\beta c_s^2\mathbf{D}+\mathbf{C})-\mathbf{M}_{1G}\big]+O(\Delta t^2).
\end{eqnarray}
Substituting Eq. (\ref{eq:4-23}) into Eq. (\ref{eq:4-22}b), one can obtain
\begin{equation}\label{eq:4-24}
\partial_t \phi+\nabla\cdot \mathbf{B} = \nabla \cdot \big[\mathbf{K}\cdot(\nabla\cdot \mathbf{D})\big] + S+\Delta t \nabla\cdot RH_1+O(\Delta t^2),
\end{equation}
where the diffusion matrix $\mathbf{K}$ is the same as that appeared in Eq. (\ref{eq:4-10}), and $RH_1$ is defined by
\begin{equation}\label{eq:4-25}
RH_1=(\mathbf{S}_1^{-1}-\mathbf{I}/2)(\partial_t \mathbf{B}+\nabla\cdot \mathbf{C})-\mathbf{S}_1^{-1}\mathbf{M}_{1G}.
\end{equation}
Under the requirement of $RH_1=0$, we can derive the following condition,
\begin{equation}\label{eq:4-26}
\mathbf{M}_{1G}=(\mathbf{I}-\mathbf{S}_1/2)(\partial_t \mathbf{B}+\nabla\cdot \mathbf{C}),
\end{equation}
which is also consistent with Eq. (\ref{eq:4-13a}). Under the condition of Eq. (\ref{eq:4-26}), we can correctly recover the NACDE (\ref{eq:4-1}) at the order of $\Delta t^2$,
\begin{equation}\label{eq:4-27}
\partial_t \phi+\nabla\cdot \mathbf{B} = \nabla \cdot [\mathbf{K}\cdot(\nabla\cdot \mathbf{D})] +S+O(\Delta t^2).
\end{equation}

We now present a special discussion on how to calculate the diffusion flux [$-\mathbf{K}\cdot(\nabla\cdot \mathbf{D})$], which is similar to the previous works \cite{Chai2013,Chai2014}. Actually, from Eqs. (\ref{eq:4-23}) and (\ref{eq:4-26}) we have
\begin{eqnarray}\label{eq:4-23a}
\nabla\cdot(\beta c_s^2\mathbf{D}) & = & - \frac{\mathbf{S}_1}{\Delta t}\mathbf{Ef}^{ne}+\big[\mathbf{M}_{1G}-(\partial_t \mathbf{B}+\nabla\cdot\mathbf{C})\big]+O(\Delta t)\nonumber\\
& = & - \frac{\mathbf{S}_1}{\Delta t}\mathbf{Ef}^{ne}+\frac{\mathbf{S}_1}{2}(\partial_t \mathbf{B}+\nabla\cdot\mathbf{C})+O(\Delta t)\nonumber\\
& = & \left\{\begin{array}{c} - \mathbf{S}_1\big(\frac{1}{\Delta t}\mathbf{Ef}^{ne}-\frac{1}{2}\mathbf{B}'S\big)+O(\Delta t),\ \ \ if\ \mathbf{B}=\mathbf{B(\phi)}\ and \ \mathbf{C}=\int\mathbf{B}'\mathbf{B}'d\phi,\\
 - \mathbf{S}_1\big(\frac{1}{\Delta t}\mathbf{Ef}^{ne}-\frac{1}{2}\partial_t \mathbf{B}\big)+O(\Delta t),\ \ \ \ \ \ \ \ \ \ \ \ \ \  \ \ \ \ \ \ \ \ \ \ \ \ \ \ \ \ \ \ \ \ \ \ \ \ \  \ if \ \mathbf{C}=\mathbf{0}.
\end{array}\right.
\end{eqnarray}

From Eq. (\ref{eq:4-23a}) we can obtain a local scheme for the diffusion flux with a second-order accuracy in time,
\begin{eqnarray}\label{eq:4-23b}
-\mathbf{K}\cdot(\nabla\cdot\mathbf{D}) = & \left\{\begin{array}{c} (\mathbf{S}_1/2-\mathbf{I})\big(\mathbf{Ef}^{ne}-\Delta t\mathbf{B}'S/2\big),\ \ if\ \mathbf{B}=\mathbf{B(\phi)}\ and \ \mathbf{C}=\int\mathbf{B}'\mathbf{B}'d\phi, \\
(\mathbf{S}_1/2-\mathbf{I})\big(\mathbf{Ef}^{ne}-\Delta t\partial_t \mathbf{B}/2\big),\ \ \ \ \ \ \ \ \ \ \ \ \ \  \ \ \ \ \ \ \ \ \ \ \ \ \ \ \ \ \ \ \ \ \ \ \ \  \ if\ \mathbf{C}=\mathbf{0},
\end{array}\right.
\end{eqnarray}
where Eq. (\ref{eq:4-10}) has been used. Additionally, if $\mathbf{D}$ is only a function of $\phi$, we can also obtain a local scheme for $\nabla\phi$ from Eq. (\ref{eq:4-23a}) or (\ref{eq:4-23b}).

\textbf{Remark}: We noted that following the idea in Ref. \cite{Shi2009}, Zhang et al. also developed a MRT model for a general isotropic CDE in a recent work \cite{Zhang2019}, and found that under the diffusive scaling ($\Delta t=\eta \Delta x^{2}$, $\eta$ is a parameter) \cite{Yoshida2010}, the general isotropic CDE can be recovered correctly from the MRT model. However, it should be noted that the diffusive scaling is an assumption, and also brings an additional constrain on the time step and lattice spacing. Besides, we would also like to point out that under the diffusive scaling, the present model and the one in Ref. \cite{Chai2016a} would reduce to the MRT model in Ref. \cite{Zhang2019}. Actually, under the diffusive scaling, the last term on the right hand side of Eq. (\ref{eq:2-1}) can be neglected since it is of fourth order of $\Delta x$, and also the terms $\partial_t \phi$ and $\partial_t \mathbf{B}$ would be absent in Eqs. (\ref{eq:4-22}a) and (\ref{eq:4-23}). In this case, the auxiliary moment $\mathbf{C}$ and first-order moment $\mathbf{M}_{1G}$ are not needed, and can be set to be zero, which also means that the auxiliary distribution function $G_{i}$ can be ignored in the evolution equation (\ref{eq:2-1}). Thus under the diffusive scaling, the evolution equation (\ref{eq:2-1}) can reduce to that in Ref. \cite{Zhang2019}.

\subsubsection{The equilibrium, auxiliary and source distribution functions of GLBM for NACDE}

As discussed above, it can be found that to recover the NACDE (\ref{eq:4-1}) correctly, some proper requirements on the equilibrium, auxiliary and source distribution functions should be satisfied. Based on Eq. (\ref{eq:4-2}) for a general D$d$Q$q$ lattice model, we can give the following expressions of $f_{j}^{eq}$, $G_j$ and $F_j$,
\begin{subequations}\label{eq:4-28}
\begin{equation}
f_j^{eq}=\omega_j\left[\phi+\frac{\textbf{c}_j\cdot{\mathbf{B}}}{c_s^2}
                   +\frac{(\beta c_s^2\mathbf{D}+\mathbf{C}-c_s^2 \phi\mathbf{I}):(\textbf{c}_j \textbf{c}_j - c_s^2\mathbf{I})}{2c_s^4}\right],
\end{equation}
\begin{equation}
G_j=\omega_j \frac{\mathbf{c}_j\cdot \mathbf{M}_{1G}}{c_s^2},
\end{equation}
\begin{equation}
F_j=\omega_j S,
\end{equation}
\end{subequations}
where $\mathbf{M}_{1G}$ is given by Eq. (\ref{eq:4-26}).

We note that when $\mathbf{D}=\phi \mathbf{I}$, the NACDE (\ref{eq:4-1}) would become
\begin{equation}\label{eq:4-29}
\partial_t \phi+\nabla\cdot\mathbf{B}= \nabla\cdot(\mathbf{K}\cdot\nabla\phi)+S.
\end{equation}
If we further take $\beta=1$ and $\mathbf{C}=0$, the equilibrium and auxiliary distribution functions $f_j^{eq}$ and $G_j$ defined by Eqs. (\ref{eq:4-28}a) and (\ref{eq:4-28}b) can be simplified as
\begin{subequations}\label{eq:4-30}
\begin{equation}
f_j^{eq}=\omega_j\big(\phi+\frac{\textbf{c}_j\cdot{\mathbf{B}}}{c_s^2}\big)
\end{equation}
\begin{equation}
G_j=\omega_j \frac{\mathbf{c}_j\cdot (\mathbf{I}-\mathbf{S}_1/2)\partial_t \mathbf{B}}{c_s^2},
\end{equation}
\end{subequations}
where the term $\partial_t \mathbf{B}$ can be computed by the first-order explicit difference scheme, i.e., $\partial_t \mathbf{B}=[\mathbf{B}(\mathbf{x}, t+\Delta t)-\mathbf{B}(\mathbf{x}, t)]/\Delta t$.

In addition, we would also like to point out that at the diffusive scaling, Eq. (\ref{eq:4-28}) can be simplified by
\begin{subequations}\label{eq:4-28a}
\begin{equation}
f_j^{eq}=\omega_j\left[\phi+\frac{\textbf{c}_j\cdot{\mathbf{B}}}{c_s^2}
                   +\frac{(\beta\mathbf{D}-\phi\mathbf{I}):(\textbf{c}_j \textbf{c}_j - c_s^2\mathbf{I})}{2c_s^2}\right],
\end{equation}
\begin{equation}
G_j=0,
\end{equation}
\begin{equation}
F_j=\omega_j S.
\end{equation}
\end{subequations}
When $\beta=1$ and $\mathbf{D}=D(\phi)\mathbf{I}$, the equilibrium distribution function (\ref{eq:4-28a}a) would reduce to the one in Ref. \cite{Zhang2019}.
\subsection{Derivation of NSEs}

We now consider the following $d$-dimensional NSEs with the source and forcing terms,
\begin{subequations}\label{eq:4-31}
\begin{equation}
\partial_t \rho+\nabla\cdot(\rho \mathbf{u})= \bar{S},
\end{equation}
\begin{equation}
\partial_t (\rho\mathbf{u})+\nabla\cdot(\rho \mathbf{uu})=-\nabla p+ \nabla\cdot \mathbf{\sigma}+ \bar{\mathbf{F}},
\end{equation}
\end{subequations}
where the shear stress $\mathbf{\sigma}$ is defined by
\begin{equation}
\mathbf{\sigma}=\mu\big[\nabla \mathbf{u}+(\nabla \mathbf{u})^T\big]+\lambda (\nabla\cdot \mathbf{u})\mathbf{I}=\mu\big[\nabla \mathbf{u}+(\nabla \mathbf{u})^T-\frac{2}{D}(\nabla\cdot \mathbf{u})\mathbf{I}\big]+\mu_{b} (\nabla\cdot \mathbf{u})\mathbf{I},
\end{equation}
where $\mu$ is the dynamic viscosity, $\lambda=\mu_{b}-2\mu/d$ with $\mu_{b}$ being the bulk viscosity \cite{Dellar2001,Kundu2016}.

Similar to the derivation of NACDE, to recover NSEs from GLBM (\ref{eq:2-1}) we also need to give some constraints on $\mathbf{\Lambda}$, $f_j$, $f_j^{eq}$, $G_j$, and $F_j$. In addition, compared to Eqs. (\ref{eq:4-2}), (\ref{eq:4-3}) and (\ref{eq:4-4}), there is another requirement on the higher-order moments of the distribution functions for NSEs. Here the following conditions should be satisfied,
\begin{subequations}\label{eq:4-32}
\begin{equation}
\rho=\sum_j f_j=\sum_j f_j^{eq},\ \ \rho \mathbf{u}=\sum_j \mathbf{c}_j f_j=\sum_j \mathbf{c}_j f_j^{eq},
\end{equation}
\begin{equation}
\sum_j \mathbf{c}_j \mathbf{c}_j f_j^{eq}= c_s^2 \rho \mathbf{I}+\rho\mathbf{uu},\ \
\sum_j \mathbf{c}_j \mathbf{c}_j \mathbf{c}_j f_j^{eq}= c_s^2 \rho \Delta \cdot \mathbf{u},
\end{equation}
\begin{equation}
\sum_j G_j=0,\ \ \sum_j \mathbf{c}_j G_j=\mathbf{M}_{1G}=0,\ \ \sum_j \mathbf{c}_j \mathbf{c}_j G_j=\mathbf{M}_{2G},
\end{equation}
\begin{equation}
\sum_j F_j=\bar{S},\ \ \sum_j \mathbf{c}_j F_j=\mathbf{M}_{1F}=\bar{\mathbf{F}},\ \ \sum_j \mathbf{c}_j \mathbf{c}_j F_j=\mathbf{M}_{2F}=0,
\end{equation}
\begin{equation}
\sum_j \mathbf{e}_{j}\mathbf{\Lambda}_{jk} = s_0\mathbf{e}_{k}, \ \ \sum_j \mathbf{c}_{j}\mathbf{\Lambda}_{jk} = \mathbf{S}_1 \mathbf{c}_k,
\end{equation}
\begin{eqnarray}
\sum_j \mathbf{c}_{j\alpha}\mathbf{c}_{j\beta}\mathbf{\Lambda}_{jk} & = & \sum_j \big(\mathbf{c}_{j\alpha}\mathbf{c}_{j\beta}-\frac{\delta_{\alpha\beta}}{d}\mathbf{c}_{j\gamma}\mathbf{c}_{j\gamma}\big)\mathbf{\Lambda}_{jk}+\sum_j \frac{\delta_{\alpha\beta}}{d}\mathbf{c}_{j\gamma}\mathbf{c}_{j\gamma}\mathbf{\Lambda}_{jk}\nonumber\\
& = &  S_{2s}\big(\mathbf{c}_{k\alpha}\mathbf{c}_{k\beta}-\frac{\delta_{\alpha\beta}}{d}\mathbf{c}_{k\gamma}\mathbf{c}_{k\gamma}\big)+S_{2b} \frac{\delta_{\alpha\beta}}{d}\mathbf{c}_{k\gamma}\mathbf{c}_{k\gamma},
\end{eqnarray}
\end{subequations}
where $\mathbf{M}_{2G}$ is a second-order tensor to be determined below, $\Delta_{\alpha\beta\gamma\theta}
=\delta_{\alpha\beta}\delta_{\gamma\theta}+\delta_{\alpha\gamma}\delta_{\beta\theta}+\delta_{\alpha\theta}\delta_{\beta\gamma}$,
$\mathbf{S}_1$ is an invertible $d\times d$ relaxation matrix, $S_{2s}$ and $S_{2b}$ are the relaxation parameters corresponding to the second-order moments.

\subsubsection{Derivation of NSEs through DTE method}

Due to the similarity between the CE analysis and and some other methods, here we only use the DTE method, i.e., Eqs. (\ref{eq:3-26}a) and (\ref{eq:3-29}), to recover NSEs. With the help of Eqs. (\ref{eq:3-31}) and (\ref{eq:3-32}), we can rewrite Eqs. (\ref{eq:3-26}a) and (\ref{eq:3-29}) as
\begin{subequations}\label{eq:4-33}
\begin{equation}
D_j f_j^{eq}=\frac{1}{\Delta t}g_j +G_j+F_j+O(\Delta t),
\end{equation}
\begin{equation}
D_j f_j^{eq}-D_j \bar{\mathbf{\Lambda}}_{jk} g_k=\frac{1}{\Delta t}g_j +G_j+F_j+\frac{\Delta t}{2}\big[(\gamma-1)\mathbf{c}_j\cdot \nabla F_j-D_j G_j\big]+O(\Delta t^2).
\end{equation}
\end{subequations}

On the other hand, from Eqs. (\ref{eq:3-31}) and (\ref{eq:4-32}) we have
\begin{subequations}\label{eq:4-34}
\begin{equation}
\sum_j \mathbf{e}_{j}g_j=-s_0\sum_k \mathbf{e}_{k}f_k^{ne}=0,\ \ \sum_j \mathbf{c}_j g_j=-\mathbf{S}_1\sum_k \mathbf{c}_k f_k^{ne}=0,
\end{equation}
\begin{equation}
\sum_j \mathbf{e}_{j}\bar{\mathbf{\Lambda}}_{jk}= (1/s_0-1/2)\mathbf{e}_{k},\ \ \sum_j \mathbf{c}_{j}\bar{\mathbf{\Lambda}}_{jk} = (\mathbf{S}_1^{-1}-\mathbf{I}/2) \mathbf{c}_k,
\end{equation}
\begin{eqnarray}
\sum_j \mathbf{c}_{j\alpha}\mathbf{c}_{j\beta}\bar{\mathbf{\Lambda}}_{jk}& = & \sum_j \big(\mathbf{c}_{j\alpha}\mathbf{c}_{j\beta}-\frac{\delta_{\alpha\beta}}{d}\mathbf{c}_{j\gamma}\mathbf{c}_{j\gamma}\big)\bar{\mathbf{\Lambda}}_{jk}+\sum_j \frac{\delta_{\alpha\beta}}{d}\mathbf{c}_{j\gamma}\mathbf{c}_{j\gamma}\bar{\mathbf{\Lambda}}_{jk}\nonumber\\
& = &  \big(\frac{1}{S_{2s}}-\frac{1}{2}\big)\big(\mathbf{c}_{k\alpha}\mathbf{c}_{k\beta}-\frac{\delta_{\alpha\beta}}{d}\mathbf{c}_{k\gamma}\mathbf{c}_{k\gamma}\big)+\big(\frac{1}{S_{2b}}-\frac{1}{2}\big) \frac{\delta_{\alpha\beta}}{d}\mathbf{c}_{k\gamma}\mathbf{c}_{k\gamma}\nonumber\\
& = &  \big(\frac{1}{S_{2s}}-\frac{1}{2}\big)\mathbf{c}_{k\alpha}\mathbf{c}_{k\beta}+\big(\frac{1}{S_{2b}}-\frac{1}{S_2s}\big) \frac{\delta_{\alpha\beta}}{d}\mathbf{c}_{k\gamma}\mathbf{c}_{k\gamma},
\end{eqnarray}
\end{subequations}
and
\begin{eqnarray}\label{eq:4-35}
\sum_j \mathbf{c}_{j\alpha} D_j \bar{\mathbf{\Lambda}}_{jk}g_k & = & \partial_t\sum_j\mathbf{c}_{j\alpha}\bar{\mathbf{\Lambda}}_{jk}g_k
+\nabla_{\beta}\cdot\sum_j\mathbf{c}_{j\alpha}\mathbf{c}_{j\beta}\bar{\mathbf{\Lambda}}_{jk}g_k\nonumber\\
& = &\partial_t (\mathbf{S}_1^{-1}-\mathbf{I}/2)\sum_k \mathbf{c}_{k\alpha} g_k+\nabla_{\beta}\cdot\sum_j\mathbf{c}_{j\alpha}\mathbf{c}_{j\beta}\bar{\mathbf{\Lambda}}_{jk}g_k\nonumber\\
& = &\nabla_{\beta}\cdot\sum_j\mathbf{c}_{j\alpha}\mathbf{c}_{j\beta}\bar{\mathbf{\Lambda}}_{jk}g_k\nonumber\\
& = & \nabla_{\beta}\cdot\sum_k\big[\big(\frac{1}{S_{2s}}-\frac{1}{2}\big)\mathbf{c}_{k\alpha}\mathbf{c}_{k\beta}+\big(\frac{1}{S_{2b}}-\frac{1}{S_{2s}}\big) \frac{\delta_{\alpha\beta}}{d}\mathbf{c}_{k\gamma}\mathbf{c}_{k\gamma}\big]g_k.
\end{eqnarray}
Summing Eq. (\ref{eq:4-33}b) and adopting Eqs. (\ref{eq:4-32}) and (\ref{eq:4-34}), we can obtain
\begin{equation}\label{eq:4-36}
\partial_t \rho+\nabla\cdot (\rho\mathbf{u}) = \frac{\Delta t}{2}(\gamma-1)\nabla\cdot \bar{\mathbf{F}}+\bar{S}+O(\Delta t^2).
\end{equation}
If we take $\gamma=1$, then the continuity equation (\ref{eq:4-31}a) is recovered correctly to the order of $O(\Delta t^2)$.

Multiplying $\mathbf{c}_j$ on both sides of Eq. (\ref{eq:4-33}), and through a summation over $j$, we have
\begin{subequations}\label{eq:4-37}
\begin{equation}
\partial_t (\rho\mathbf{u})+\nabla\cdot (c_s^2\rho \mathbf{I}+\rho\mathbf{uu}) = \bar{\mathbf{F}}+O(\Delta t),
\end{equation}
\begin{eqnarray}
\partial_t (\rho\mathbf{u}_{\alpha})+\nabla_{\beta}\cdot (c_s^2\rho \delta_{\alpha\beta}+\rho\mathbf{u}_{\alpha}\mathbf{u}_{\beta}) & - & \nabla_{\beta}\cdot \sum_k\big[\big(\frac{1}{S_{2s}}-\frac{1}{2}\big)\mathbf{c}_{k\alpha}\mathbf{c}_{k\beta}+\big(\frac{1}{S_{2b}}-\frac{1}{S_{2s}}\big) \frac{\delta_{\alpha\beta}}{d}\mathbf{c}_{k\gamma}\mathbf{c}_{k\gamma}\big]g_k\nonumber \\
& = & \bar{\mathbf{F}}_{\alpha}-\frac{\Delta t}{2}\nabla_{\beta}\cdot \mathbf{M}_{2G,\alpha\beta}+O(\Delta t^2),
\end{eqnarray}
\end{subequations}
where Eqs. (\ref{eq:4-32}), (\ref{eq:4-34}) and (\ref{eq:4-35}) have been used.

Additionally, from Eqs. (\ref{eq:4-32}) and (\ref{eq:4-33}) we get
\begin{eqnarray}\label{eq:4-38}
\sum_k \mathbf{c}_{k\alpha}\mathbf{c}_{k\beta} g_k & = &\Delta t \sum_k \mathbf{c}_{k\alpha}\mathbf{c}_{k\beta} (D_k f_k^{eq}-G_k-F_k)+O(\Delta t^2)\nonumber\\
& = &\Delta t\big(\partial_t \sum_k \mathbf{c}_{k\alpha}\mathbf{c}_{k\beta} f_k^{eq}+\nabla_{\gamma}\cdot \sum_k \mathbf{c}_{k\alpha}\mathbf{c}_{k\beta}\mathbf{c}_{k\gamma} f_k^{eq}-\mathbf{M}_{2G,\alpha\beta}\big)+O(\Delta t^2)\nonumber\\
& = &\Delta t \big[\partial_t (c_s^2\rho \delta_{\alpha\beta}+\rho\mathbf{u_{\alpha}u_{\beta}})+\nabla_{\gamma}\cdot(c_s^2 \rho \Delta_{\alpha\beta\gamma\theta} \cdot \mathbf{u_{\theta}})-\mathbf{M}_{2G,\alpha\beta}\big]+O(\Delta t^2)\nonumber\\
& = &\Delta t \big[\partial_t (c_s^2\rho \delta_{\alpha\beta}+\rho\mathbf{u_{\alpha}u_{\beta}})+\nabla_{\gamma}\cdot c_s^2 \rho (\mathbf{u_{\alpha}}\delta_{\beta\gamma}+ \mathbf{u_{\beta}}\delta_{\alpha\gamma}+\mathbf{u_{\gamma}}\delta_{\alpha\beta})-\mathbf{M}_{2G,\alpha\beta}\big]\nonumber\\
& + &O(\Delta t^2).
\end{eqnarray}

Based on the following equations,
\begin{subequations}
\begin{equation}\label{eq:4-45}
\partial_t (\rho\mathbf{uu})=\mathbf{u}\bar{\mathbf{F}}+\bar{\mathbf{F}}\mathbf{u}-c_s^2\big[\mathbf{u}\nabla\rho+(\mathbf{u}\nabla\rho)^T\big]
-\nabla\cdot(\rho\mathbf{uuu})-\mathbf{uu}\bar{S}+O(\Delta t),
\end{equation}
\begin{eqnarray}\label{eq:4-40}
\nabla\cdot(c_s^2 \rho \mathbf{u})+\nabla\cdot(c_s^2 \rho \mathbf{u})^T
=c_s^2 \rho\big[\nabla \mathbf{u}+(\nabla \mathbf{u})^T\big]+c_s^2\big[\mathbf{u}\nabla\rho+(\mathbf{u}\nabla\rho)^T\big],
\end{eqnarray}
\end{subequations}
we can rewrite Eq. (\ref{eq:4-38}) as
\begin{subequations}\label{eq:4-38a}
\begin{eqnarray}
\sum_k \mathbf{c}_{k\alpha}\mathbf{c}_{k\beta} g_k = \Delta t \big[\rho c_{s}^{2}(\partial_{\alpha}\mathbf{u}_{\beta}+\partial_{\beta}\mathbf{u}_{\alpha})+\mathbf{u}_{\alpha}\bar{\mathbf{F}}_{\beta}+\bar{\mathbf{F}}_{\alpha}\mathbf{u}_{\beta}+(c_{s}^{2}\delta_{\alpha\beta}-\mathbf{u}_{\alpha}\mathbf{u}_{\beta})\bar{S}
-\mathbf{M}_{2G,\alpha\beta}\big]+ O(\Delta t^2),
\end{eqnarray}
\begin{eqnarray}
\sum_k \mathbf{c}_{k\gamma}\mathbf{c}_{k\gamma} g_k \delta_{\alpha\beta} = \Delta t \big[2\rho c_{s}^{2}\partial_{\gamma}\mathbf{u}_{\gamma}+2\mathbf{u}_{\gamma}\bar{\mathbf{F}}_{\gamma}+(dc_{s}^{2}-\mathbf{u}_{\gamma}\mathbf{u}_{\gamma})\bar{S}
-\mathbf{M}_{2G,\gamma\gamma}\big]\delta_{\alpha\beta}+ O(\Delta t^2),
\end{eqnarray}
\end{subequations}
where the term $\nabla_{\gamma}\cdot(\rho\mathbf{u}_{\alpha}\mathbf{u}_{\beta}\mathbf{u}_{\gamma})$ have been neglected, this is because for incompressible fluid flows, it is the order of $O(Ma^{3})$ with $Ma$ being the Mach number.
Substituting Eq. (\ref{eq:4-38a}) into Eq. (\ref{eq:4-37}b) and using Eq. (\ref{eq:4-36}), we can obtain
\begin{eqnarray}\label{eq:4-41}
& & \partial_t (\rho\mathbf{u}_{\alpha}) + \nabla_{\beta}\cdot (c_s^2\rho \delta_{\alpha\beta} + \rho\mathbf{u}_{\alpha}\mathbf{u}_{\beta})\nonumber\\
& = &\nabla_{\beta}\cdot \big[\Delta t \big(\frac{1}{S_{2s}}-\frac{1}{2}\big) c_s^2\rho\big(\nabla_{\alpha}\mathbf{u}_{\beta}+\nabla_{\beta}\mathbf{u}_{\alpha}-\frac{2}{d}\mathbf{u}_{\gamma}\delta_{\alpha\beta}\big)+\frac{2}{d}\Delta t\big(\frac{1}{S_{2b}}-\frac{1}{2}\big) c_s^2\rho\nabla_{\gamma}\mathbf{u}_{\gamma}\delta_{\alpha\beta}\big]+ \bar{\mathbf{F}}_{\alpha}\nonumber\\
& + &\Delta t \nabla_{\beta}\cdot (RH_2)_{\alpha\beta}+O(\Delta t^2),
\end{eqnarray}
where
\begin{eqnarray}\label{eq:4-42}
(RH_2)_{\alpha\beta}&=& \big(\frac{1}{S_{2s}}-\frac{1}{2}\big)\big[\bar{\mathbf{F}}_{\alpha}\mathbf{u}_{\beta}+\mathbf{u}_{\alpha}\bar{\mathbf{F}}_{\beta}-\frac{2}{d}\mathbf{u}_{\gamma}\bar{\mathbf{F}}_{\gamma}\delta_{\alpha\beta}
-\big(\mathbf{u}_{\alpha}\mathbf{u}_{\beta}-\frac{1}{d}\mathbf{u}_{\gamma}^{2}\delta_{\alpha\beta}\big)\bar{S}-\mathbf{M}_{2G,\alpha\beta}+\frac{1}{d}\mathbf{M}_{2G,\gamma\gamma}\delta_{\alpha\beta}\big]\nonumber \\
& + & \big(\frac{1}{S_{2b}}-\frac{1}{2}\big)\big[\frac{2}{d}\mathbf{u}_{\gamma}\bar{\mathbf{F}}_{\gamma}\delta_{\alpha\beta}+\big(c_{s}^{2}-\frac{1}{d}\mathbf{u}_{\gamma}^{2}\big)\delta_{\alpha\beta}\bar{S}-\frac{1}{d}\mathbf{M}_{2G,\gamma\gamma}\delta_{\alpha\beta}
\big]-\frac{\Delta t}{2}\mathbf{M}_{2G,\alpha\beta}\nonumber\\
&=& \big(\frac{1}{S_{2s}}-\frac{1}{2}\big)\big[\bar{\mathbf{F}}_{\alpha}\mathbf{u}_{\beta}+\mathbf{u}_{\alpha}\bar{\mathbf{F}}_{\beta}
-\big(\mathbf{u}_{\alpha}\mathbf{u}_{\beta}-c_{s}^{2}\delta_{\alpha\beta}\big)\bar{S}-\mathbf{M}_{2G,\alpha\beta}\big]\nonumber \\
& + & \big(\frac{1}{S_{2b}}-\frac{1}{S_{2s}}\big)\big[\frac{2}{d}\mathbf{u}_{\gamma}\bar{\mathbf{F}}_{\gamma}+\big(c_{s}^{2}-\frac{1}{d}\mathbf{u}_{\gamma}^{2}\big)\bar{S}-\frac{1}{d}\mathbf{M}_{2G,\gamma\gamma}
\big]\delta_{\alpha\beta}-\frac{\Delta t}{2}\mathbf{M}_{2G,\alpha\beta}.
\end{eqnarray}
To obtain the correct NSEs, $(RH_2)_{\alpha\beta}=0$, which gives rise to following equations,
\begin{subequations}\label{eq:4-46}
\begin{equation}\label{eq:4-46a}
\mathbf{M}_{2G,\gamma\gamma}=\big(1-\frac{S_{2b}}{2}\big)\big[2\mathbf{u}_{\gamma}\bar{\mathbf{F}}_{\gamma}+(dc_s^2-\mathbf{u}_{\gamma}\mathbf{u}_{\gamma})\bar{S}\big],
\end{equation}
\begin{eqnarray}\label{eq:4-46b}
\mathbf{M}_{2G,\alpha\beta} & = &\big(1-\frac{S_{2s}}{2}\big)\big[\bar{\mathbf{F}}_{\alpha}\mathbf{u}_{\beta}+\bar{\mathbf{F}}_{\beta}\mathbf{u}_{\alpha}
+(c_{s}^{2}\delta_{\alpha\beta}-\mathbf{u}_{\alpha}\mathbf{u}_{\beta})\bar{S}-\frac{2}{d}\mathbf{u}_{\gamma}\bar{\mathbf{F}}_{\gamma}-\big(c_{s}^{2}-\frac{\mathbf{u}_{\gamma}\mathbf{u}_{\gamma}}{d}\big)\bar{S}\delta_{\alpha\beta}\big]\nonumber\\
& + & \big(1-\frac{S_{2b}}{2}\big)\big[\frac{2}{d}\mathbf{u}_{\gamma}\bar{\mathbf{F}}_{\gamma}+\big(c_{s}^{2}-\frac{\mathbf{u}_{\gamma}\mathbf{u}_{\gamma}}{d}\big)\bar{S}\delta_{\alpha\beta}\big].
\end{eqnarray}
\end{subequations}
Introduce the matrix $\mathbf{W}=\bar{\mathbf{F}}\mathbf{u}+\mathbf{u}\bar{\mathbf{F}}
+(c_{s}^{2}\mathbf{I}-\mathbf{u}\mathbf{u})\bar{S}$, then Eq. (\ref{eq:4-46b}) can be written as
\begin{eqnarray}\label{eq:4-46bb}
\mathbf{M}_{2G} & = &\big(1-\frac{S_{2s}}{2}\big)\big[\mathbf{W}-\textbf{tr}(\mathbf{W})\big]+\big(1-\frac{S_{2b}}{2}\big)\textbf{tr}(\mathbf{W}),
\end{eqnarray}
where $\textbf{tr}(\mathbf{W})$ is the trace of the matrix $\mathbf{W}$. We note that when $\bar{S}=0$ and $S_{2s}=S_{2b}$, Eq. (\ref{eq:4-46bb}) would be the same as the second-order moment of the forcing term distribution function in the previous work \cite{Guo2002}.

Based on $\mathbf{M}_{2G}$ defined by Eq. (\ref{eq:4-46bb}), we can write Eq. (\ref{eq:4-41}) in the following form,
\begin{eqnarray}\label{eq:4-44}
\partial_t (\rho\mathbf{u})+\nabla\cdot (c_s^2\rho \mathbf{I}+\rho\mathbf{uu})=\nabla\cdot\mu\big[\nabla\mathbf{u}+(\nabla\mathbf{u})^{T}-\frac{2}{d}(\nabla\cdot\mathbf{u})\mathbf{I}\big]+\nabla(\mu_{b}\nabla\cdot\mathbf{u})+ \bar{\mathbf{F}}+O(\Delta t^2).
\end{eqnarray}
When $p=\rho c_s^2$ and the truncation error $O(\Delta t^2)$ is neglected, Eq. (\ref{eq:4-44}) would reduce to the momentum equation (\ref{eq:4-31}b) with the following dynamic and bulk viscosities,
\begin{equation}\label{eq:4-45}
\mu= \big(\frac{1}{S_{2s}}-\frac{1}{2}\big) \rho c_s^2\Delta t,\ \mu_{b}= \frac{2}{d}\big(\frac{1}{S_{2b}}-\frac{1}{2}\big) \rho c_s^2\Delta t.
\end{equation}

Finally, we also present a local scheme to calculate shear stress or strain rate tensor in the framework of LBM \cite{Chai2012,Kruger2010,Yong2012}. To this end, from Eq. (\ref{eq:4-38a}a) we can first obtain the following expression of strain rate tensor with a first-order accuracy in time,
\begin{eqnarray} \label{eq:4-45a}
\frac{\partial_{\alpha}\mathbf{u}_{\beta}+\partial_{\beta}\mathbf{u}_{\alpha}}{2} &= &
\frac{1}{2\rho c_{s}^{2}\Delta t} \sum_k \mathbf{c}_{k\alpha}\mathbf{c}_{k\beta} g_k- \frac{1}{2\rho c_{s}^{2}}\big[\mathbf{u}_{\alpha}\bar{\mathbf{F}}_{\beta}+\bar{\mathbf{F}}_{\alpha}\mathbf{u}_{\beta}+(c_{s}^{2}\delta_{\alpha\beta}-\mathbf{u}_{\alpha}\mathbf{u}_{\beta})\bar{S}
-\mathbf{M}_{2G,\alpha\beta}\big]\nonumber\\
&=& \frac{1}{2\rho c_{s}^{2}\Delta t}\sum_k \big[-S_{2s}\mathbf{c}_{k\alpha}\mathbf{c}_{k\beta}+(S_{2b}-S_{2s})\frac{\delta_{\alpha\beta}}{d}\mathbf{c}_{k\gamma}\mathbf{c}_{k\gamma}\big] f_k^{ne}\nonumber\\
& -&  \frac{1}{2\rho c_{s}^{2}}\big[\mathbf{u}_{\alpha}\bar{\mathbf{F}}_{\beta}+\bar{\mathbf{F}}_{\alpha}\mathbf{u}_{\beta}+(c_{s}^{2}\delta_{\alpha\beta}-\mathbf{u}_{\alpha}\mathbf{u}_{\beta})\bar{S}
-\mathbf{M}_{2G,\alpha\beta}\big].
\end{eqnarray}
Then based on Eq. (\ref{eq:4-45a}), one can also give a local scheme for shear stress with a second-order accuracy in time,
\begin{eqnarray} \label{eq:4-45b}
\mu (\partial_{\alpha}\mathbf{u}_{\beta}+\partial_{\beta}\mathbf{u}_{\alpha})
& = &\big(\frac{1}{S_{2s}}-
\frac{1}{2}\big) \sum_k \big[-S_{2s}\mathbf{c}_{k\alpha}\mathbf{c}_{k\beta}+(S_{2b}-S_{2s})\frac{\delta_{\alpha\beta}}{d}\mathbf{c}_{k\gamma}\mathbf{c}_{k\gamma}\big] f_k^{ne}\nonumber\\
& - & \big(\frac{1}{S_{2s}}-
\frac{1}{2}\big)\Delta t\big[\mathbf{u}_{\alpha}\bar{\mathbf{F}}_{\beta}+\bar{\mathbf{F}}_{\alpha}\mathbf{u}_{\beta}+(c_{s}^{2}\delta_{\alpha\beta}-\mathbf{u}_{\alpha}\mathbf{u}_{\beta})\bar{S}
-\mathbf{M}_{2G,\alpha\beta}\big].
\end{eqnarray}
We note that Eqs. (\ref{eq:4-45a}) and (\ref{eq:4-45b}) are the same as the results in Ref. \cite{Chai2012} once $\bar{S}=0$.

\subsubsection{The equilibrium, auxiliary and source distribution functions of GLBM for NSEs}

From above discussion, one can clearly observe that to recover the macroscopic NSEs (\ref{eq:4-31}) from GLBM (\ref{eq:2-1}), the equilibrium, auxiliary and source distribution functions should satisfy some necessary requirements, as depicted by Eq. (\ref{eq:4-32}). For a general D$d$Q$q$ lattice model, the explicit expressions of $f_{j}^{eq}$, $G_j$ and $F_j$ can be given by
\begin{subequations}\label{eq:4-46}
\begin{equation}
f_j^{eq}=\omega_j\rho\left[1+\frac{\textbf{c}_j\cdot{\mathbf{u}}}{c_s^2}
                   +\frac{\mathbf{uu}:(\textbf{c}_j \textbf{c}_j - c_s^2\mathbf{I})}{2c_s^4}\right],
\end{equation}
\begin{equation}
G_j=\omega_j \frac{ \mathbf{M}_{2G}:(\textbf{c}_j \textbf{c}_j - c_s^2\mathbf{I})}{c_s^2},
\end{equation}
\begin{equation}
F_j=\omega_j \big(\bar{S}+\frac{\textbf{c}_j\cdot{\bar{\mathbf{F}}}}{c_s^2}\big),
\end{equation}
\end{subequations}
where $\mathbf{M}_{2G}$ is determined by Eq. (\ref{eq:4-46bb}).

\section{Collision matrix and some special cases of GLBM}

When the discrete velocity set $V_q=\{\mathbf{c}_j, 0\leq j\leq q-1\}$ is given, we can construct different forms of collision matrix $\mathbf{\Lambda}$ in GLBM. In general, the collision matrix $\mathbf{\Lambda}$ in the commonly used MRT model has the following form \cite{Guo2013,Kruger2017},
\begin{equation}\label{eq:5-1}
\mathbf{\Lambda}=\mathbf{M}^{-1}\mathbf{SM},\ \  \mathbf{S}=\textbf{diag}(\mathbf{S}_0,\ \mathbf{S}_1,\ \mathbf{S}_2,\ \cdots,\ \mathbf{S}_m),
\end{equation}
where the rows of transformation matrix $\mathbf{M}$ are made up of discrete velocities in $V_q$, $\mathbf{S}$ is a block diagonal matrix, $\mathbf{S}_k \in R^{n_k\times n_k}$ is a relaxation matrix corresponding to the $k$-th ($0\leq k\leq m$) order moment of discrete velocity space. Denote $\mathbf{M}$ with the following form,
\begin{equation}\label{eq:5-2}
\mathbf{M}=(\mathbf{M}_0^T,\ \mathbf{M}_1^T,\ \mathbf{M}_2^T,\ \cdots,\ \mathbf{M}_m^T)^T, \ \mathbf{M}_k \in R^{n_k\times q},\ 0\leq k\leq m,
\end{equation}
then we have
\begin{equation}\label{eq:5-3}
\mathbf{M}_k\mathbf{\Lambda}=\mathbf{S}_k \mathbf{M}_k,\  0\leq k\leq m,
\end{equation}
with
\begin{equation}\label{eq:5-4}
S_0 \in R,\ \mathbf{S}_1 \in R^{d\times d},\ \mathbf{S}_2  \in R^{\bar{d}\times \bar{d}},\ \bar{d}=d(d+1)/2.
\end{equation}

For the GLBM (1) discussed above, we only need the first two left-eigenvectors of the collision matrix for NACDE,
\begin{equation}\label{eq:5-3}
\mathbf{M}_0=\mathbf{e}=(1, 1, \cdots , 1),\ \mathbf{M}_1=\mathbf{E}=(\mathbf{c}_0, \mathbf{c}_1, \cdots , \mathbf{c}_{q-1}),
\end{equation}
and the other eigenvectors can be chosen arbitrarily. However, for the Navier-Stokes equations, only the first three left-eigenvectors are needed,
\begin{equation}\label{eq:5-3}
\mathbf{M}_0=\mathbf{e},\ \mathbf{M}_1=\mathbf{E}, \ \mathbf{M}_2=\left(\begin{array}{c}
\mathbf{M}_{2}^{(1)}  \\
\mathbf{M}_{2}^{(2)}
\end{array} \right),
\end{equation}
where $\mathbf{M}_{2}^{(1)}=(c_{0\alpha}c_{0\alpha}, c_{1\alpha}c_{1\alpha}, \cdots , c_{q-1,\alpha}c_{q-1,\alpha})$ and $\mathbf{M}_{2}^{(2)}=(c_{0\alpha}c_{0\beta}, c_{1\alpha}c_{1\beta}, \cdots , c_{q-1,\alpha}c_{q-1,\beta})_{\alpha<\beta}$. Based on Eq. (\ref{eq:4-32}f), we can determine the relaxation matrix $\mathbf{S}_{2}$ as
\begin{equation}\label{eq:5-5}
\mathbf{S}_{2}=S_{2s}\mathbf{I}+\frac{2(S_{2b}-S_{2s})}{d}\left(\begin{array}{cc}
\mathbb{E}_{d} & 0 \\
0 & 0
\end{array} \right)_{\bar{d}},
\end{equation}
where $\mathbb{E}_{d}$ is a $d\times d$ matrix with all elements $(\mathbb{E}_{d})_{ij}=1$. Here it should also be noted that for the orthogonal eigenvectors, we can obtain the similar results as those in the previous section, and refer the reader to Ref. \cite{Kaehler2013} for some details.

We now point out that some commonly used LB models in the literature can be obtained from our GLBM when the matrices $\mathbf{\Lambda}$, $\mathbf{M}$ and $\mathbf{S}$ are specified with some special forms.

\begin{enumerate}[(1)]
\item Lattice BGK or SRT model \cite{Qian1992,Chen1992}:
\begin{equation}\label{eq:5-5}
\mathbf{\Lambda}=\mathbf{S}=\frac{1}{\tau}\mathbf{I}.
\end{equation}
\item TRT model \cite{Ginzburg2005a,Ginzburg2008}:
\begin{equation}\label{eq:5-6}
\mathbf{S}_{2k}=s^{+}\mathbf{I}_{2k},\ \mathbf{S}_{2k+1}=s^{-}\mathbf{I}_{2k+1},\ 0\leq 2k, 2k+1\leq m.
\end{equation}
Let
\begin{subequations}\label{eq:5-7}
\begin{equation}
q-1=2\bar{m},
\end{equation}
\begin{equation}
\mathbf{V}_q=\{\mathbf{c}_j,\ j=0, 1, \cdots, 2\bar{m}\}=\{\mathbf{c}_0,\mathbf{c}_1, \cdots , \mathbf{c}_{\bar{m}}, -\mathbf{c}_1, \cdots , -\mathbf{c}_{\bar{m}}\},
\end{equation}
\end{subequations}
the collision matrix $\mathbf{\Lambda}$ can be written as
\begin{equation}\label{eq:5-8}
\mathbf{\Lambda}=
\left[
\begin{array}{ccc}
\omega+\bar{\omega} &   0    &0\\
    0  & \omega \mathbf{I}_{\bar{m}} & \bar{\omega} \mathbf{I}_{\bar{m}}\\
    0  & \bar{\omega} \mathbf{I}_{\bar{m}} & \omega \mathbf{I}_{\bar{m}}
\end{array}
\right],
\end{equation}
where $s^{+}+s^{-}=2\omega, s^{+}-s^{-}=2\bar{\omega}$, or equivalently, $s^{+}=\omega+\bar{\omega}, s^{-}=\omega-\bar{\omega}$ with $s^{+}$ and $s^{-}$ being the relaxation parameters corresponding to the symmetric and anti-symmetric parts of distribution functions.

In addition, we also note that if $\mathbf{c}_0=\mathbf{0}$ is not contained in the discrete velocity set $\mathbf{V}_q$, the matrix $\mathbf{\Lambda}$ in Eq. (\ref{eq:5-8}) would be given by
\begin{equation}\label{eq:5-9}
\mathbf{\Lambda}=
\left[
\begin{array}{cc}
     \omega \mathbf{I}_{\bar{m}} & \bar{\omega} \mathbf{I}_{\bar{m}}\\
     \bar{\omega} \mathbf{I}_{\bar{m}} & \omega \mathbf{I}_{\bar{m}}
\end{array}
\right],
\end{equation}

\item RLB models \cite{Latt2006,Chen2006,WangL2015,WangL2018}: For simplicity, we first introduce the following tensors,
\begin{eqnarray}\label{eq:5-8}
\mathbf{R}_{jk}=\omega_j \frac{\mathbf{c}_j\cdot\mathbf{c}_k}{c_s^2},\ \mathbf{P}_{jk}=\omega_j\frac{\mathbf{Q}_j:\mathbf{c}_k\mathbf{c}_k}{2c_s^4},\ \mathbf{Q}_j=\mathbf{c}_j\mathbf{c}_j-c_s^2\mathbf{I}.
\end{eqnarray}
 In the RLB model, the collision matrix $\mathbf{\Lambda}$ is taken as $\mathbf{\Lambda}=\mathbf{I}-(1-1/\tau)\mathbf{R}$ for NACDE, while for NSEs, $\mathbf{\Lambda}=\mathbf{I}-(1-1/\tau)\mathbf{P}$. It is clear that as a general form, we can take $\mathbf{\Lambda}=\mathbf{I}-(1-1/\tau)(\mathbf{R+P})$ for both NACDE and NSEs.

\item MLK schemes \cite{Yang2014,Wang2015}: For the NACDE (\ref{eq:4-1}), the collision matrix $\mathbf{\Lambda}$ in the MLK scheme is given by
\begin{equation}\label{eq:5-7}
\mathbf{\Lambda}=\frac{1}{\tau}\mathbf{I}+\big(\frac{1}{\tau-A}-\frac{1}{\tau}\big)\mathbf{R},
\end{equation}
while for the NSEs (\ref{eq:4-31}), it should be taken as
\begin{equation}
\mathbf{\Lambda}=\frac{1}{\tau}\mathbf{I}+\big(\frac{1}{\tau-A}-\frac{1}{\tau}\big)\mathbf{P}.
\end{equation}
As a general case, the collision matrix $\mathbf{\Lambda}$ can be specified by
\begin{equation}
\mathbf{\Lambda}=\frac{1}{\tau}\mathbf{I}+\big(\frac{1}{\tau-A}-\frac{1}{\tau}\big)(\mathbf{R}+\mathbf{P}),
\end{equation}
where $\mathbf{R}$ and $\mathbf{P}$ are the same as those in Eq. (\ref{eq:5-8}).

\item Classical MRT models \cite{dHumieres1992,Lallemand2000,dHumieres2002}: In the classical MRT models, the collision matrix $\mathbf{\Lambda}$ can be determined by
\begin{equation}
\mathbf{\Lambda}=\mathbf{M}^{-1}\mathbf{SM},
\end{equation}
where $\mathbf{S}=\textbf{diag}(S_0,\ S_1,\ S_2,\ \cdots,\ S_{q-1})$ is a standard diagonal relaxation matrix, $\mathbf{M}$ is the transformation matrix and is composed of the orthogonal eigenvectors \cite{Guo2013,Kruger2017}.

\item Block triple-relaxation-time LB model \cite{Zhao2019}: For brevity we first introduce the following $\bar{\mathbf{R}}$ and $\bar{\mathbf{P}}$ in the block triple-relaxation-time LB model,
\begin{eqnarray}\label{eq:5-8}
\bar{\mathbf{R}}_{jk}=\omega_j \frac{\mathbf{c}_j\cdot\big[(\mathbf{S}_{1}-S_{0}\mathbf{I})\mathbf{c}_k\big]}{c_s^2},\ \bar{\mathbf{P}}_{jk}=\omega_j\frac{\mathbf{Q}_j:\big[(\mathbf{K}_{2}-S_{0}\hat{\mathbf{I}})\circ(\mathbf{c}_k\mathbf{c}_k)\big]}{2c_s^4},
\end{eqnarray}
where $\hat{\mathbf{I}}$ is a matrix with $\hat{\mathbf{I}}_{ij} = 1$, the symbol $\circ$ represents the Hadamard product, $\mathbf{K}_{2}$ is a tensor related to $\mathbf{S}_{2}$ \cite{Zhao2019}. Then the collision matrix $\mathbf{\Lambda}$ can be given by
\begin{equation}
\mathbf{\Lambda}=S_{0}\mathbf{I}+\bar{\mathbf{R}}+\bar{\mathbf{P}}.
\end{equation}
\end{enumerate}

\section{Conclusions}

In this work, we developed a general framework for the modeling of the GLBM for the NACDE and NSEs, and also presented a detailed analysis on the C-E analysis, MI, DTE and RE approaches that have been adopted to derive the macroscopic equations from the LB models. The results show that mathematically, these four analysis methods are equivalent to each other, and can obtain the same macroscopic governing equations from present GLBM. In addition, we provided some details on the equilibrium, auxiliary and source distribution functions, and some special discussion on how to develop the local schemes for the diffusion flux (and/or $\nabla\phi$) and shear stress (and/or stain rate tensor). Finally, we also pointed out that the existing LB models, including the lattice BGK or SRT model, TRT model, RLB model, MLK Schemes, the classical MRT model and the block triple-relaxation-time LB model, are some special cases of the present GLBM.

\section*{Acknowledgements}
This work was financially supported by the National Natural Science
Foundation of China (Grant Nos. 51836003 and 51576079) and the
National Key Research and Development Program of China (Grant No.
2017YFE0100100).

\section{Appendix}

\subsection{Appendix A: From discrete-velocity Boltzmann equation to lattice Boltzmann equation}

The GLBM [Eq. (\ref{eq:2-1})] can be obtained from the discrete-velocity Boltzmann equation (DVBE) with invertible collision matrix $\tilde{\mathbf{\Lambda}}$,
\begin{equation}\label{eq:A1}
\partial_t f_j(\mathbf{x},t)+\mathbf{c}_j\cdot \nabla f_j(\mathbf{x},t)=-\tilde{\mathbf{\Lambda}}_{jk}f_k^{ne}(\mathbf{x},t)+\tilde{G}_j(\mathbf{x},t)+F_j(\mathbf{x},t),
\end{equation}
which can be also considered as a basic equation to develop some other mesoscopic methods (e.g., the discrete-unified gas-kinetic scheme \cite{Guo2013a, Wu2016}).

Compared to Eq. (\ref{eq:2-1}), the source distribution function in DVBE (\ref{eq:A1}) is the sum of $F_j$ and $\tilde{G}_j$, $F_j$ is used as the source or forcing term, $\tilde{G}_j$ is adopted to remove some additional terms.

Integrating Eq. (\ref{eq:A1}) along the characteristic line $\mathbf{x}'=\mathbf{x}+\mathbf{c}_j t'$ with $t' \in [0, \Delta t]$, we obtain
\begin{equation}\label{eq:A2}
f_j(\mathbf{x}+\mathbf{c}_j \Delta t,t+\Delta t)\\
=f_j(\mathbf{x},t)\nonumber+\int_0^{\Delta t}\big(-\tilde{\mathbf{\Lambda}}_{jk}f_k^{ne}+\tilde{G}_j+F_j\big)(\mathbf{x}+\mathbf{c}_j t',t+ t')dt',
\end{equation}
Based on the following results,
\begin{subequations}\label{eq:A3}
\begin{eqnarray}
\int_0^{\Delta t}\big(-\tilde{\mathbf{\Lambda}}_{jk}f_k^{ne}+\tilde{G}_j\big)(\mathbf{x}+\mathbf{c}_j t',t+ t')dt'
& = & \frac{\Delta t}{2}\big[\big(-\tilde{\mathbf{\Lambda}}_{jk}f_k^{ne}+\tilde{G}_j\big)(\mathbf{x}+\mathbf{c}_j \Delta t,t+\Delta t)+\big(-\tilde{\mathbf{\Lambda}}_{jk}f_k^{ne}
+\tilde{G}_j\big)(\mathbf{x},t)\big]\nonumber\\
& + &O(\Delta t^3),
\end{eqnarray}
\begin{eqnarray}
\int_0^{\Delta t}F_j(\mathbf{x}+\mathbf{c}_j t',t+ t')dt' & = & \int_0^{\Delta t}[F_j(\mathbf{x},t)+t'D_j F_j(\mathbf{x},t)+O(t'^2)]dt'\nonumber\\
& = &\Delta t \big[F_j(\mathbf{x},t)+\frac{\Delta t}{2}D_j F_j(\mathbf{x},t)\big]+O(\Delta t^3),
\end{eqnarray}
\end{subequations}
we can write Eq. (\ref{eq:A2}) as
\begin{equation}\label{eq:A4}
\bar{f}_j(\mathbf{x}+\mathbf{c}_j \Delta t,t+\Delta t)= \tilde{f}_j(\mathbf{x},t)+O(\Delta t^3),
\end{equation}
where
\begin{subequations}\label{eq:A5}
\begin{equation}
\bar{f}_j=f_j-\frac{\Delta t}{2}\big(-\tilde{\mathbf{\Lambda}}_{jk}f_k^{ne}+\tilde{G}_j\big),
\end{equation}
\begin{equation}
\tilde{f}_j=f_j+\frac{\Delta t}{2}\big(-\tilde{\mathbf{\Lambda}}_{jk}f_k^{ne}+\tilde{G}_j\big)+\Delta t\big(F_j+\frac{\Delta t}{2}D_j F_j\big).
\end{equation}
\end{subequations}
Actually, Eq. (\ref{eq:A5}) can also be written as
\begin{subequations}\label{eq:A6}
\begin{equation}
\bar{\mathbf{f}}=\mathbf{f}-\frac{\Delta t}{2}\big(-\tilde{\mathbf{\Lambda}}\mathbf{f}^{ne}+\tilde{\mathbf{G}}\big)=\big(\mathbf{I}+\frac{\Delta t}{2}\tilde{\mathbf{\Lambda}}\big)\mathbf{f}-\frac{\Delta t}{2}\big(\tilde{\mathbf{\Lambda}} \mathbf{f}^{eq}+\tilde{\mathbf{G}}\big),
\end{equation}
\begin{equation}
\tilde{\mathbf{f}}=\big(\mathbf{I}-\frac{\Delta t}{2}\tilde{\mathbf{\Lambda}}\big)\mathbf{f}+\frac{\Delta t}{2}(\tilde{\mathbf{\Lambda}} \mathbf{f}^{eq}+\tilde{\mathbf{G}})+\Delta t\big(\mathbf{F}+\frac{\Delta t}{2}\mathbf{DF}\big).
\end{equation}
\end{subequations}
From above equation, we have
\begin{subequations}\label{eq:A7}
\begin{equation}
\mathbf{f}=\big(\mathbf{I}+\frac{\Delta t}{2}\tilde{\mathbf{\Lambda}}\big)^{-1}\big[\bar{\mathbf{f}}+\frac{\Delta t}{2}(\tilde{\mathbf{\Lambda}} \mathbf{f}^{eq}+\tilde{\mathbf{G}})\big],
\end{equation}
\begin{equation}
\tilde{\mathbf{f}}=\big(\mathbf{I}-\frac{\Delta t}{2}\tilde{\mathbf{\Lambda}}\big)(\mathbf{I}+\frac{\Delta t}{2}\tilde{\mathbf{\Lambda}})^{-1}\big[\bar{\mathbf{f}}+\frac{\Delta t}{2}(\tilde{\mathbf{\Lambda}} \mathbf{f}^{eq}+\tilde{\mathbf{G}})\big]
+\frac{\Delta t}{2}(\tilde{\mathbf{\Lambda}} \mathbf{f}^{eq}+\tilde{\mathbf{G}})+\Delta t\big(\mathbf{F}+\frac{\Delta t}{2}\mathbf{DF}\big).
\end{equation}
\end{subequations}
Let
\begin{equation}\label{eq:A8}
\mathbf{I}-\mathbf{\Lambda}=\big(\mathbf{I}-\frac{\Delta t}{2}\tilde{\mathbf{\Lambda}}\big)\big(\mathbf{I}+\frac{\Delta t}{2}\tilde{\mathbf{\Lambda}}\big)^{-1},
\end{equation}
then
\begin{equation}\label{eq:A9}
\tilde{\mathbf{f}}=(\mathbf{I}-\mathbf{\Lambda})\bar{\mathbf{f}}+\big[\mathbf{I}+(\mathbf{I}-\mathbf{\Lambda})\big]\big[\frac{\Delta t}{2}(\tilde{\mathbf{\Lambda}} \mathbf{f}^{eq}+\tilde{\mathbf{G}})\big]+\Delta t\big[\mathbf{F}+\frac{\Delta t}{2}\mathbf{DF}\big].
\end{equation}
According to the following equation,
\begin{eqnarray}\label{eq:A10}
\big[\mathbf{I}+(\mathbf{I}-\mathbf{\Lambda})\big]\frac{\Delta t}{2}\tilde{\mathbf{\Lambda}} & = & \big[\mathbf{I}+\big(\mathbf{I}-\frac{\Delta t}{2}\tilde{\mathbf{\Lambda}}\big)\big(\mathbf{I}+\frac{\Delta t}{2}\tilde{\mathbf{\Lambda}}\big)^{-1}\big]\frac{\Delta t}{2}\tilde{\mathbf{\Lambda}}\nonumber\\
& = &\frac{\Delta t}{2}\tilde{\mathbf{\Lambda}}+\big(\mathbf{I}-\frac{\Delta t}{2}\tilde{\mathbf{\Lambda}}\big)\big(\mathbf{I}+\frac{\Delta t}{2}\tilde{\mathbf{\Lambda}}\big)^{-1}\big(\mathbf{I}+\frac{\Delta t}{2}\tilde{\mathbf{\Lambda}}-\mathbf{I}\big)\nonumber\\
& = &\frac{\Delta t}{2}\tilde{\mathbf{\Lambda}}+\big(\mathbf{I}-\frac{\Delta t}{2}\tilde{\mathbf{\Lambda}}\big)-\big(\mathbf{I}-\frac{\Delta t}{2}\tilde{\mathbf{\Lambda}}\big)\big(\mathbf{I}+\frac{\Delta t}{2}\tilde{\mathbf{\Lambda}}\big)^{-1}=\mathbf{\Lambda},
\end{eqnarray}
where Eq. (\ref{eq:A8}) has been used, we can write Eq. (\ref{eq:A9}) as
\begin{eqnarray}\label{eq:A11}
\tilde{\mathbf{f}}  & = & (\mathbf{I}-\mathbf{\Lambda})\bar{\mathbf{f}}+\mathbf{\Lambda} \mathbf{f}^{eq}+\Delta t\big(\mathbf{I}-\frac{1}{2}\mathbf{\Lambda}\big)\tilde{\mathbf{G}}+\Delta t\big(\mathbf{F}+\frac{\Delta t}{2}\mathbf{DF}\big)\nonumber\\
& = & \bar{\mathbf{f}}-\mathbf{\Lambda} (\bar{\mathbf{f}}-\mathbf{f}^{eq})+\Delta t\big[\big(\mathbf{I}-\frac{1}{2}\mathbf{\Lambda}\big)\tilde{\mathbf{G}}+\mathbf{F}+\frac{\Delta t}{2}\mathbf{DF}\big].
\end{eqnarray}
Substituting above equation into Eq. (\ref{eq:A4}) and removing the term $O(\Delta t^3)$, we obtain
\begin{eqnarray}\label{eq:A12}
\bar{f}_j(\mathbf{x}+\mathbf{c}_j \Delta t,t+\Delta t)=\bar{f}_j-\mathbf{\Lambda}_{jk} (\bar{f}_k-f_k^{eq})+\Delta t\big(G_j+F_j+\frac{\Delta t}{2}D_jF_j\big),
\end{eqnarray}
where $\mathbf{G}=\big(\mathbf{I}-\mathbf{\Lambda}/2\big)\tilde{\mathbf{G}}$, the relaxation matrix $\mathbf{\Lambda}$ is given by
\begin{eqnarray}\label{eq:A13}
\mathbf{\Lambda} & = & \mathbf{I}-\big(\mathbf{I}-\frac{\Delta t}{2}\tilde{\mathbf{\Lambda}}\big)\big(\mathbf{I}+\frac{\Delta t}{2}\tilde{\mathbf{\Lambda}}\big)^{-1}\nonumber\\
& = & \big(\mathbf{I}+\frac{\Delta t}{2}\tilde{\mathbf{\Lambda}}\big)\big(\mathbf{I}+\frac{\Delta t}{2}\tilde{\mathbf{\Lambda}}\big)^{-1}-\big(\mathbf{I}-\frac{\Delta t}{2}\tilde{\mathbf{\Lambda}}\big)\big(\mathbf{I}+\frac{\Delta t}{2}\tilde{\mathbf{\Lambda}}\big)^{-1}\nonumber\\
& = &\Delta t\tilde{\mathbf{\Lambda}}\big(\mathbf{I}+\frac{\Delta t}{2}\tilde{\mathbf{\Lambda}}\big)^{-1}\nonumber\\
& = &\big(\frac{1}{2}\mathbf{I}+\frac{1}{\Delta t}\tilde{\mathbf{\Lambda}}^{-1}\big)^{-1}.
\end{eqnarray}
We note that Eq. (\ref{eq:A12}) has the same form as Eq. (\ref{eq:2-1}) except the last term on the right side. Actually if $\gamma=1$, Eq. (\ref{eq:2-1}) would reduce to Eq. (\ref{eq:A12}), while $\gamma=0$ can also be used derive correct NACDE \cite{Chai2016a}.

\subsection{Appendix B: From discrete-velocity Boltzmann equation to the macroscopic equations through CEM}

It is known from above discussion that two classes of macroscopic equations, i.e., NACDE and NSEs, can be recovered correctly from GLBM (\ref{eq:2-1}) through some analysis methods (CE analysis, MI, DTE and RE methods) when the moments of the distribution functions $f_j^{eq}$, $G_j$ and $F_j$ are given properly. Here only the CE analysis is considered to obtain the NACDE and NSEs.

Based on the CE analysis, we have
\begin{subequations}\label{eq:B1}
\begin{equation}
f_j=f_j=f_j^{(0)}+\epsilon
f_j^{(1)}+\epsilon^2 f_j^{(2)},\ \tilde{G}_j=\epsilon \tilde{G}_j^{(1)}+\epsilon^2 \tilde{G}_j^{(2)},\ F_j=\epsilon F_j^{(1)}+\epsilon^2 F_j^{(2)},
\end{equation}
\begin{equation}
\partial_t=\epsilon\partial_{t_1}+\epsilon^2\partial_{t_2},\ \nabla=\epsilon\nabla_1,\ D_j=\epsilon D_{1j}+\epsilon^2\partial_{t_2}.
\end{equation}
\end{subequations}
Substituting Eq. (\ref{eq:B1}) into Eq. (\ref{eq:A1}), we can obtain the following equations at the order of $\epsilon^0$, $\epsilon^1$ and $\epsilon^2$,
\begin{subequations}\label{eq:B2}
\begin{equation}
f_j^{(0)}-f_j^{eq}=0,
\end{equation}
\begin{equation}
D_{1j} f_j^{(0)} = - \tilde{\mathbf{\Lambda}}_{jl}f_l^{(1)}+\tilde{G}_j^{(1)} +F_j^{(1)},
\end{equation}
\begin{equation}
\partial_{t_2} f_j^{(0)}+ D_{1j} f_j^{(1)}
 = -\tilde{\mathbf{\Lambda}}_{jl} f_l^{(2)}+\tilde{G}_j^{(2)} +F_j^{(2)}.
\end{equation}
\end{subequations}

It can be seen that Eq. (\ref{eq:B2}) derived from DVBE (\ref{eq:A1}) is simpler than Eq. (\ref{eq:3-3}) obtained from GLBM (1) in which some discrete terms [e.g., $\Delta t^{2}D_{1j}f_{j}^{(0)}/2$ in Eq. (\ref{eq:3-3}c)] caused by the numerical scheme are included.

\subsubsection{From DVBE to NACDE through CE analysis}

To correctly recover NACDE (\ref{eq:4-1}) from the DVBE (\ref{eq:A1}), the unknown conserved scalar $\phi$, $f_j$ and $f_j^{eq}$ satisfy Eq. (\ref{eq:4-2}), while moments of $\tilde{G}_j$, $F_j$ and $\tilde{\Lambda}$ are given by
\begin{subequations}\label{eq:B3}
\begin{equation}
\sum_j \tilde{G}_j=0,\ \sum_j \mathbf{c}_j \tilde{G}_j=\mathbf{M}_{1\tilde{G}},
\end{equation}
\begin{equation}
\sum_j F_j=S,\ \sum_j \mathbf{c}_j F_j=\mathbf{M}_{1F}=0,
\end{equation}
\begin{equation}
\sum_j \mathbf{e}_{j}\tilde{\mathbf{\Lambda}}_{jk} = \tilde{s}_0\mathbf{e}_{k},\ \sum_j \mathbf{c}_{j}\tilde{\mathbf{\Lambda}}_{jk} = \tilde{\mathbf{S}}_1 \mathbf{c}_k,\ \forall k,
\end{equation}
\end{subequations}
where the moment $\mathbf{M}_{1\tilde{G}}$ to be determined below. $\tilde{\mathbf{S}}_1$ is an invertible $d\times d$ relaxation matrix corresponding to the diffusion matrix $\mathbf{K}$.

Similar to Eq. (\ref{eq:4-4}), it follows from Eqs. (\ref{eq:4-2}), (\ref{eq:B1}), (\ref{eq:B2}a) and (\ref{eq:B3}) that
\begin{subequations}\label{eq:B4}
\begin{equation}
\sum_j f_j^{(k)}=0,
\end{equation}
\begin{equation}
\sum_j \tilde{G}_j^{(k)}=0,\ \sum_j \mathbf{c}_j \tilde{G}_j^{(k)}=\mathbf{M}_{1\tilde{G}}^{(k)},
\end{equation}
\begin{equation}
\sum_j F_j^{(k)}=S^{(k)},\ \sum_j \mathbf{c}_j F_j^{(k)}=\mathbf{0},
\end{equation}
\end{subequations}
where $k\geq 1$. Summing Eq. (\ref{eq:B2}b) and Eq. (\ref{eq:B2}c) over $j$, one can obtain
\begin{subequations}\label{eq:B5}
\begin{equation}
\partial_{t_1} \phi +  \nabla_1\cdot\mathbf{B}= -\tilde{s}_0\sum_l f_l^{(1)}+ S^{(1)}=S^{(1)},
\end{equation}
\begin{equation}
\partial_{t_2} \phi + \nabla_1\cdot \sum_l \mathbf{c}_l f_l^{(1)} = S^{(2)}.
\end{equation}
\end{subequations}
where Eqs. (\ref{eq:4-2}), (\ref{eq:B3}) and (\ref{eq:B4}) have been used.

Multiplying $\mathbf{c}_j$ on both sides of Eqs. (\ref{eq:B2}b) and (\ref{eq:B2}c), and summing them over $j$, we have
\begin{subequations}\label{eq:B6}
\begin{equation}
\partial_{t_1} \mathbf{B} +  \nabla_1\cdot(\mathbf{C}+\beta c_s^2 \mathbf{D})= -\tilde{\mathbf{S}}_1\sum_l \mathbf{c}_l f_l^{(1)}+ \mathbf{M}_{1\tilde{G}}^{(1)},
\end{equation}
\begin{equation}
\partial_{t_2} \mathbf{B} +\partial_{t_1}\sum_l \mathbf{c}_l f_l^{(1)}+ \nabla_1\cdot \sum_l \mathbf{c}_l \mathbf{c}_l f_l^{(1)}
= -\tilde{\mathbf{S}}_1\sum_l \mathbf{c}_l f_l^{(2)}+\mathbf{M}_{1\tilde{G}}^{(2)},
\end{equation}
\end{subequations}
where Eqs. (\ref{eq:B3}) and (\ref{eq:B4}) have been adopted to derive above equation.

Substituting Eq. (\ref{eq:B6}a) into Eq. (\ref{eq:B5}b) gives
\begin{equation}\label{eq:B8}
\partial_{t_2} \phi = \nabla_1\cdot [\mathbf{K}\cdot(\nabla_1\cdot \mathbf{D})]+\nabla_1 \cdot RH_1 + S^{(2)},
\end{equation}
where $\mathbf{K}=\beta c_s^2 \tilde{\mathbf{S}}_1^{-1}$ with $\tilde{\mathbf{S}}_1^{-1}=(\mathbf{S}_1^{-1}-\frac{1}{2}\mathbf{I})\Delta t$, $RH_1=\tilde{\mathbf{S}}_1^{-1}\big(\partial_{t_1} \mathbf{B}+\nabla_1 \cdot \mathbf{C}-\mathbf{M}_{1\tilde{G}}^{(1)}\big)$.
Taking $\mathbf{M}_{1\tilde{G}}$ such that
\begin{equation}\label{eq:B10}
\partial_t \mathbf{B}+\nabla \cdot \mathbf{C}-\mathbf{M}_{1\tilde{G}}=0,
\end{equation}
we have $RH_1=0$, which leads to
\begin{equation}\label{eq:B11}
\partial_{t_2} \phi = \nabla_1\cdot [\mathbf{K}\cdot(\nabla_1\cdot \mathbf{D})]+ S^{(2)}.
\end{equation}
Combining Eq. (\ref{eq:B6}a) with Eq. (\ref{eq:B11}) yields the NACDE (\ref{eq:4-1}),
\begin{equation}\label{eq:B12}
\partial_{t} \phi + \nabla \cdot \mathbf{B} = \nabla\cdot [\mathbf{K}\cdot(\nabla\cdot \mathbf{D})]+ S.
\end{equation}

In addition, it should be noted that from Eq. (\ref{eq:B10}) and the relation $\mathbf{G}=\big(\mathbf{I}-\mathbf{\Lambda}/2\big)\tilde{\mathbf{G}}$, we can determine $\mathbf{M}_{1G}$ as
\begin{equation}\label{eq:B13}
 \mathbf{M}_{1G}=(\mathbf{I}-\mathbf{S}_1/2)\mathbf{M}_{1\tilde{G}}=(\mathbf{I}-\mathbf{S}_1/2)(\partial_t \mathbf{B}+\nabla \cdot \mathbf{C}),
\end{equation}
which is the same as Eq. (\ref{eq:4-13a}).

\subsubsection{From DVBE to NSEs through CEM}

Similar to above discussion, to recover NSEs (\ref{eq:4-31}) from the DVBE (\ref{eq:A1}), the macroscopic variables $\rho$ and $\mathbf{u}$, and the distribution functions $f_j$ and $f_j^{eq}$ satisfy Eqs. (\ref{eq:4-32}a) and (\ref{eq:4-32}b), while the moments of $\tilde{G}_j$, $F_j$ and $\tilde{\mathbf{\Lambda}}$ should satisfy the following relations,
\begin{subequations}\label{eq:B14}
\begin{equation}
\sum_j \tilde{G}_j=0,\ \sum_j \mathbf{c}_j \tilde{G}_j=\mathbf{M}_{1\tilde{G}}=\mathbf{0},\ \sum_j \mathbf{c}_j \mathbf{c}_j \tilde{G}_j=\mathbf{M}_{2\tilde{G}},
\end{equation}
\begin{equation}
\sum_j F_j=\bar{S},\ \sum_j \mathbf{c}_j F_j=\mathbf{M}_{1F}=\bar{\mathbf{F}}, \ \sum_j \mathbf{c}_j \mathbf{c}_j F_j=\mathbf{M}_{2F}=0,
\end{equation}
\begin{equation}
\sum_j \mathbf{e}_{j}\tilde{\mathbf{\Lambda}}_{jk} = \tilde{s}_0 \mathbf{e}_{k},\ \sum_j \mathbf{c}_{j}\tilde{\mathbf{\Lambda}}_{jk} = \tilde{\mathbf{S}}_1 \mathbf{c}_k,
\end{equation}
\begin{eqnarray}
\sum_j \mathbf{c}_{j\alpha}\mathbf{c}_{j\beta}\tilde{\mathbf{\Lambda}}_{jk} & = & \sum_j \big(\mathbf{c}_{j\alpha}\mathbf{c}_{j\beta}-\frac{\delta_{\alpha\beta}}{d}\mathbf{c}_{j\gamma}\mathbf{c}_{j\gamma}\big)\tilde{\mathbf{\Lambda}}_{jk}+\sum_j \frac{\delta_{\alpha\beta}}{d}\mathbf{c}_{j\gamma}\mathbf{c}_{j\gamma}\tilde{\mathbf{\Lambda}}_{jk}\nonumber\\
& = &  \tilde{S}_{2s}\big(\mathbf{c}_{k\alpha}\mathbf{c}_{k\beta}-\frac{\delta_{\alpha\beta}}{d}\mathbf{c}_{k\gamma}\mathbf{c}_{k\gamma}\big)+\tilde{S}_{2b} \frac{\delta_{\alpha\beta}}{d}\mathbf{c}_{k\gamma}\mathbf{c}_{k\gamma},
\end{eqnarray}
\end{subequations}
where the moment $\mathbf{M}_{2\tilde{G}}$ will be determined later. $\tilde{S}_{2s}$ and $\tilde{S}_{2b}$ are two relaxation parameters related to dynamic viscosity $\mu$ and bulk viscosity $\mu_{b}$.

Similarly, from Eqs. (\ref{eq:4-32}), (\ref{eq:B1}), (\ref{eq:B2}a) and (\ref{eq:B14}) we have
\begin{subequations}\label{eq:B15}
\begin{equation}
\sum_j f_j^{(k)}=0,\ \sum_j \mathbf{c}_j f_j^{(k)}=0,
\end{equation}
\begin{equation}
\sum_j \tilde{G}_j^{(k)}=0,\ \sum_j \mathbf{c}_j \tilde{G}_j^{(k)}=\mathbf{0},\ \sum_j \mathbf{c}_j \mathbf{c}_j \tilde{G}_j^{(k)}=\mathbf{M}_{2\tilde{G}}^{(k)},
\end{equation}
\begin{equation}
\sum_j F_j^{(k)}=\bar{S}^{(k)}, \sum_j \mathbf{c}_j F_j^{(k)}=\bar{\mathbf{F}}^{(k)}, \sum_j \mathbf{c}_j \mathbf{c}_j F_j^{(k)}=\mathbf{0},
\end{equation}
\end{subequations}
where $k\geq 1$.
Summing Eqs. (\ref{eq:B2}b) and (\ref{eq:B2}c) over $j$, and using Eqs. (\ref{eq:4-32}a), (\ref{eq:B14}) and (\ref{eq:B15}), we can obtain
\begin{subequations}\label{eq:B16}
\begin{equation}
\partial_{t_1} \rho +  \nabla_1\cdot (\rho\mathbf{u})= -\tilde{s}_0\sum_l f_l^{(1)}+ \bar{S}^{(1)}=\bar{S}^{(1)},
\end{equation}
\begin{equation}
\partial_{t_2} \rho  = -\tilde{s}_0\sum_l f_l^{(2)}+ \bar{S}^{(2)}=\bar{S}^{(2)}.
\end{equation}
\end{subequations}
Multiplying $\mathbf{c}_j$ and $\mathbf{c}_j\mathbf{c}_j$ on both sides of Eqs. (\ref{eq:B2}b) and (\ref{eq:B2}c), and summing these equations over $j$, one can obtain
\begin{subequations}\label{eq:B17}
\begin{equation}
\partial_{t_1} (\rho\mathbf{u}) +  \nabla_1\cdot( c_s^2 \rho\mathbf{I}+\rho\mathbf{uu})= -\tilde{\mathbf{S}}_1\sum_l \mathbf{c}_l f_l^{(1)}+ \mathbf{M}_{1F}^{(1)}=\bar{\mathbf{F}}^{(1)},
\end{equation}
\begin{equation}
\partial_{t_2} (\rho\mathbf{u}) + \nabla_1\cdot \sum_l \mathbf{c}_l \mathbf{c}_l f_l^{(1)}= -\tilde{\mathbf{S}}_1\sum_l \mathbf{c}_l f_l^{(2)}+\mathbf{M}_{1F}^{(2)}=\bar{\mathbf{F}}^{(2)},
\end{equation}
\end{subequations}

\begin{subequations}\label{eq:B18}
\begin{equation}
\partial_{t_1} ( c_s^2 \rho\mathbf{I}+\rho\mathbf{uu}) +  \nabla_1\cdot[ c_s^2 \Delta \cdot(\rho\mathbf{u}) ]= -\tilde{S}_{2s}\sum_l \big(\mathbf{c}_l\mathbf{c}_l -\frac{1}{d}\mathbf{c}_{l}\cdot\mathbf{c}_{l}\mathbf{I}\big)f_l^{(1)}-\tilde{S}_{2b} \sum_l\big(\frac{1}{d}\mathbf{c}_{l}\cdot\mathbf{c}_{l}\mathbf{I}\big)f_l^{(1)}+\mathbf{M}_{2\tilde{G}}^{(1)},
\end{equation}
\begin{eqnarray}
\partial_{t_2} ( c_s^2 \rho\mathbf{I}+\rho\mathbf{uu}) +\partial_{t_1}\sum_l \mathbf{c}_l\mathbf{c}_l f_l^{(1)}+ \nabla_1\cdot \sum_l \mathbf{c}_l\mathbf{c}_l \mathbf{c}_l f_l^{(1)} \nonumber\\
= -\tilde{S}_{2s}\sum_l \big(\mathbf{c}_l\mathbf{c}_l -\frac{1}{d}\mathbf{c}_{l}\cdot\mathbf{c}_{l}\mathbf{I}\big)f_l^{(2)}-\tilde{S}_{2b} \sum_l\big(\frac{1}{d}\mathbf{c}_{l}\cdot\mathbf{c}_{l}\mathbf{I}\big)f_l^{(2)}+\mathbf{M}_{2\tilde{G}}^{(2)},
\end{eqnarray}
\end{subequations}
where Eqs. (\ref{eq:B14}) and (\ref{eq:B15}) have been used.

From Eq. (\ref{eq:B16}), the continuity equation (\ref{eq:4-31}a) is recovered correctly to the order of $O(\epsilon ^2)$,
\begin{equation}\label{eq:B18aa}
\partial_{t} \rho +  \nabla\cdot (\rho\mathbf{u})= \bar{S}.
\end{equation}
On the other hand, with the help of Eq. (\ref{eq:B18}), we can obtain the momentum equation (\ref{eq:4-31}b) from Eq. (\ref{eq:B17}).
According to Eqs. (\ref{eq:B14}a) and (\ref{eq:B17}a), we can rewrite Eq. (\ref{eq:B18}a) as
\begin{eqnarray}\label{eq:B20}
\partial_{t_1} ( c_s^2 \rho\mathbf{I}+\rho\mathbf{uu}) & + & \nabla_1\cdot[ c_s^2 \Delta \cdot(\rho\mathbf{u}) ] = c_s^2 \bar{S}^{(1)}\mathbf{I}+\mathbf{u}\bar{\mathbf{F}}^{(1)}+\bar{\mathbf{F}}^{(1)}\mathbf{u} + \mathbf{uu}\bar{S}^{(1)} + \rho c_s^2\big[\nabla_1\mathbf{u}+(\nabla_1\mathbf{u})^T\big]\nonumber\\
&= & -\tilde{S}_{2s}\sum_l \big(\mathbf{c}_l\mathbf{c}_l -\frac{1}{d}\mathbf{c}_{l}\cdot\mathbf{c}_{l}\mathbf{I}\big)f_l^{(1)}-\tilde{S}_{2b} \sum_l\big(\frac{1}{d}\mathbf{c}_{l}\cdot\mathbf{c}_{l}\mathbf{I}\big)f_l^{(1)}+\mathbf{M}_{2\tilde{G}}^{(1)}.
\end{eqnarray}
To derive correct momentum equation, the following conditions should be satisfied,
\begin{subequations}\label{eq:B21}
\begin{equation}
\rho c_s^2\big[\nabla_1\mathbf{u}+(\nabla_1\mathbf{u})^T\big]=-\tilde{S}_{2s}\sum_l \big(\mathbf{c}_l\mathbf{c}_l -\frac{1}{d}\mathbf{c}_{l}\cdot\mathbf{c}_{l}\mathbf{I}\big)f_l^{(1)}-\tilde{S}_{2b} \sum_l\big(\frac{1}{d}\mathbf{c}_{l}\cdot\mathbf{c}_{l}\mathbf{I}\big)f_l^{(1)},
\end{equation}
\begin{equation}
\mathbf{M}_{2\tilde{G}}^{(1)}=c_s^2 \bar{S}^{(1)}\mathbf{I}+\mathbf{u}\bar{\mathbf{F}}^{(1)}+\bar{\mathbf{F}}^{(1)}\mathbf{u} + \mathbf{uu}\bar{S}^{(1)}.
\end{equation}
\end{subequations}
Then from Eq. (\ref{eq:B21}), we can obtain
\begin{equation}\label{eq:B22}
\sum_{l}\mathbf{c}_l\mathbf{c}_lf_l^{(1)}=-\rho c_s^2\big\{\frac{1}{\tilde{S}_{2s}}\big[\nabla_1\mathbf{u}+(\nabla_1\mathbf{u})^T-\frac{2}{d}\big(\nabla_1\cdot\mathbf{u}\big)\mathbf{I}\big]+\frac{1}{\tilde{S}_{2b}}\frac{2}{d}\big(\nabla_1\cdot\mathbf{u}\big)\mathbf{I}\big\}.
\end{equation}
Substituting Eqs. (\ref{eq:B22}) into Eq. (\ref{eq:B17}b) yields the following equation,
\begin{equation}\label{eq:B23}
\partial_{t_2} (\rho\mathbf{u})=\rho c_s^2\big\{\frac{1}{\tilde{S}_{2s}}\big[\nabla_1\mathbf{u}+(\nabla_1\mathbf{u})^T-\frac{2}{d}\big(\nabla_1\cdot\mathbf{u}\big)\mathbf{I}\big]+\frac{1}{\tilde{S}_{2b}}\frac{2}{d}\big(\nabla_1\cdot\mathbf{u}\big)\mathbf{I}\big\}+\bar{\mathbf{F}}^{(2)}.
\end{equation}
Combining Eq. (\ref{eq:B17}a) with Eq. (\ref{eq:B23}) we have
\begin{equation}\label{eq:B24}
\partial_t (\rho\mathbf{u})+\nabla\cdot (\rho\mathbf{uu})=-\nabla P+\nabla\cdot\mu\big[\nabla\mathbf{u}+(\nabla\mathbf{u})^{T}-\frac{2}{d}(\nabla\cdot\mathbf{u})\mathbf{I}\big]+\nabla(\mu_{b}\nabla\cdot\mathbf{u})+ \bar{\mathbf{F}},
\end{equation}
where the pressure $P$, the dynamic and bulk viscosities are given by
\begin{equation}\label{eq:B25}
P=\rho c_{s}^{2},\ \mu=\rho c_s^2/\tilde{S}_{2s},\ \mu_{b}=2\rho c_s^2/( \tilde{S}_{2b}d).
\end{equation}
From Eqs. (\ref{eq:B18aa}) and (\ref{eq:B24}), it is clear that through the CE analysis, the NSEs (\ref{eq:4-31}) can be correctly recovered from the DVBE (\ref{eq:A1}) at the order of $O(\epsilon ^2)$. In addition, we would also like to point out that the relaxation parameters $\tilde{S}_{2s}$ and $\tilde{S}_{2b}$ in the DVBE are related to $S_{2s}$ and $S_{2b}$ in the GLBM through $\tilde{S}_{2s}^{-1}=(S_{2s}^{-1}-1/2)\Delta t$ and $\tilde{S}_{2b}^{-1}=(S_{2b}^{-1}-1/2)\Delta t$.

\end{document}